\documentclass[11pt]{article}
\usepackage[margin=2.5cm]{geometry}
\usepackage{xcolor}
\usepackage{pdfpages}
\usepackage{amsmath}
\usepackage{hyperref}
\usepackage{graphicx}
\usepackage{listings}
\usepackage{amssymb}
\usepackage{hyperref}
\usepackage[hypcap=true]{caption}
\usepackage{caption}
\usepackage{subcaption}

\usepackage{datetime}
\newdate{date}{01}{05}{2013}
\date{\displaydate{date}}

\numberwithin{equation}{section}

\begin{document}

\title{\textcolor{black}{{\begin{huge}An Introduction to Fluid Dynamics and Numerical 
Solution of Shock Tube Problem by Using Roe Solver\end{huge}}}}

\author{\begin{Large}Soumen Roy                       \end{Large} \\ \begin{large}St.Xavier's College \& Bose Institute
 \end{large}\\ \begin{large} Department of Physics \end{large}\\ \begin{large}Kolkata 700091 \end{large}}

\maketitle

\date
\begin{figure}[h!]
 \centering
 \includegraphics[width=5 cm,height=5 cm,bb = 0 0 197 257,keepaspectratio=true]{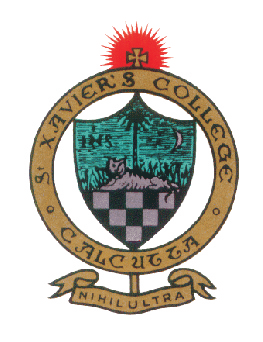}
\end{figure}
\vspace{7 cm}

\thispagestyle{empty}
\pagenumbering{roman}

\newpage
\renewcommand{\abstractname}{Acknowledgements}
\begin{abstract}
\noindent First and foremost, I offer my sincerest gratitude to my supervisor,
Prof. Sanjay K. Ghosh, who has supported me throughout my thesis with his
patience and knowledge whilst allowing me the room to work in my own way.
I attribute the level of my Masters degree to his encouragement and effort
and without him this thesis, too, would not have been completed or written.
One simply could not wish for a better or friendlier supervisor.

\vspace{1 mm}

\noindent I would like to thank the administration of Bose Institute, which played an 
instrumental role during the course of my project by providing me requisite
computer facility.

\vspace{1 mm}

\noindent My parents, Sukumar and Lila Roy, have been a constant source of
support – emotional, moral and of source financial – during my whole life.
\end{abstract}

\newpage

\renewcommand{\abstractname}{Abstract}
\begin{abstract}
\vspace{5 mm}
The study of fluids has been a topic of intense research for several hundred years. Over the years, this has further increased due to improved computational facility, which makes it easy to numerically simulate the fluid dynamics, which was beyond the analytical reach before. In this project, we discuss the general transport theorem for moving control volume system, apply this theory to control mass system which gives the continuity equation, and on momentum conservation from which we get the Navier-Stokes equation and energy conservation. By approximating the three equations for the ideal gas flow, we get one dimensional Euler equation. These equations are non-linear, and their analytic solutions are highly non-trivial. We will use Riemann solvers to get the exact solution numerically. In this project, we have considered only the Sod's Shock Tube problem. Finally, we focus on the exact and approximate solution of density, pressure, velocity, entropy, and Mach number using Python code.
\end{abstract}

\newpage
\tableofcontents

\addtocontents{section}

\newpage

\setcounter{page}{1}
\pagenumbering{arabic}
\section{Introduction}

\subsection{What is a Fluid? }
\noindent A fluid is any substance that deforms continuously when 
subjected to a shear stress, no matter how small. Usually a liquid or a gas.

\subsection{Newton's Law of Viscosity:}
\noindent The shear viscosity of a fluid expresses its resistance to shearing flows,
 where adjacent layers move parallel to each other with different speeds. It can be 
defined through the idealized situation known as a Couette flow, where a layer of fluid
 is trapped between two horizontal plates, one fixed and one moving horizontally at 
constant speed . (The plates are assumed to be very large, so that one need not
 consider what happens near their edges.)If the speed of the top plate is small enough, 
the fluid particles will move parallel to it, and their speed will vary linearly 
from zero at the bottom to  at the top. Each layer of fluid will move faster than
 the one just below it, and friction between them will give rise to a force
 resisting their relative motion. In particular, the fluid will apply on the 
top plate a force in the direction opposite to its motion, and an equal but 
opposite to the bottom plate. An external force is therefore required in order
 to keep the top plate moving at constant speed.

\begin{figure}[h]
 \centering
 \includegraphics[width=12cm,height=12cm,bb=0 0 896 595,keepaspectratio=true]{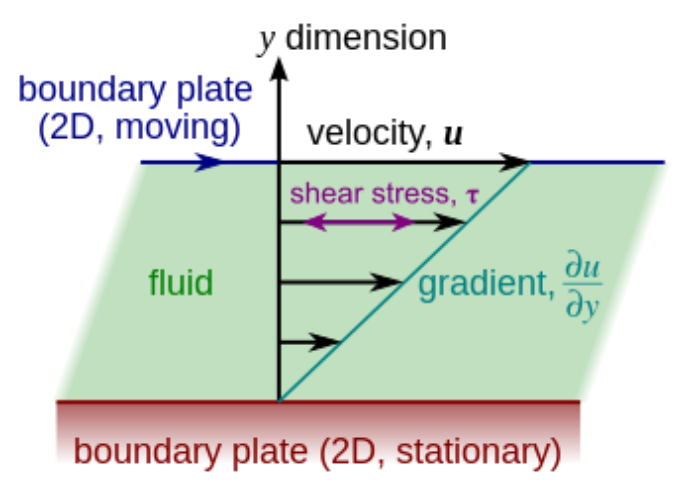}
 \caption{Laminar shear of fluid between two plates.}
 \label{fig:1}
\end{figure}

\vspace{3 mm}

\noindent The magnitude $\textbf{F}$ of this force is found to be proportional 
to the speed $\textbf{u}$ and the area A of each plate , and inversely
 proportional to the separation y. i.e.

\[\textbf{F} = \mu A \frac{\textbf{u}}{y}\]

\noindent the proportionality factor $\mu$ is this formula the viscosity of
 the fluid The ratio $\frac{\textbf{u}}{y}$  is called the rate of shear deformation or 
shear velocity, and is the derivative of the fluid speed in the
 direction perpendicular to the plates. Isaac Newton expressed 
the viscous forces by the differential equation

\[\tau = \mu \frac{\partial \textbf{u}}{\partial y}\]

\noindent where $\tau = \frac{\textbf{f}}{A}$ and $\frac{\partial \textbf{u}}{\partial y}$
is the local shear velocity. and similar case for 2 D

\[\tau_{xy} = \mu\left(\frac{\partial \textbf{u}}{\partial y} + \frac{\partial \textbf{v}}{\partial x}\right)\]

\noindent This is also known as $\textbf{Newton's Second law of Viscosity}$.

\subsection{Specification of motion}
Fluids are treated as continuous media. There are four main physical idea
which forms the basis of fluid dynamics:
\begin{enumerate}
 \item[$\ast$] The continuum hypothesis,
\item[$\ast$] Conservation of mass,
\item[$\ast$] Balance of momentum (Newton‘s second law of motion) and
\item[$\ast$] Balance (conservation) of energy. The last of these is not needed in the description of some types of flows
\end{enumerate}

\noindent Motion and state
can be specified in terms of the velocity $\textbf{u}$, pressure  P,
density  $\rho$ etc evaluated at every point in space x 
and time t. To define the density at a point, for example, suppose
the point to be surrounded by a very small element (small compared
with length scales of interest in experiments) which nevertheless
contains a very large number of molecules. The density is then the 
total mass of all the molecules in the element divided by the volume
of the element. Considering the velocity, pressure etc as functions 
of time and position in space is consistent with measurement techniques
using fixed instruments in moving fluids. It is called the Eulerian
specification. However, Newton's laws of motion (see below) are
expressed in terms of individual particles, or fluid elements, which move 
about. Specifying a fluid motion in terms of the position $\textbf{x}(t)$
of an individual particle (identified by its initial position, say) is 
called the  Lagrangian specification. The two are linked by the fact that
the velocity of such an element is equal to the velocity of the fluid
evaluated at the position occupied by the element:

\vspace{3 mm}

\[\frac{d\textbf{x}}{dt} = \textbf{u}(\textbf{x}(t),t)\]

\vspace{3 mm}

\noindent $\textbf{Path Line:}$ The path followed by a fluid element is called a particle path or
path line. Mathematically obtained by

\[d\textbf{x} = \textbf{u} dt\]
\[\Rightarrow \textbf{x} = \textbf{x}_{0} + \int_{t_{0}}^{t_{1}}\textbf{u}(x(t),t)dt\]

\noindent $\textbf{Stream Line:}$ A streamline is a continuous line within a fluid such that the tangent at each point
is the direction of the velocity vector at that point.stream lines are defined as 

\[\frac{d\textbf{x}}{dt}\otimes\textbf{u} = 0\]

\vspace{3 mm}

\noindent the components of $\textbf{u}$ and $\textbf{x}$ are related by from upper expression

\[\frac{dx}{u} = \frac{dy}{v} = \frac{dz}{dw}\]

\noindent $\textbf{Streak Lines:}$ A streak line is the locus of the temporary 
locations of all particles that have passed though a fixed point in the flow
field at any instant of time.
 
\subsection{Material Derivative}

\noindent Identify (or label) a material of the fluid; track (or follow) it as it moves, and monitor change in its
properties. The properties may be velocity, temperature, density, mass, concentration etc in the
flow field.
\vspace{3 mm}
In $\textbf{Lagrangian Approach}$ the trajectory of each individual fluid parcel as it
moves from some initial location, often described as ``―placing a
coordinate system on each fluid parcel`` and ``―riding on that parcel as it
travels through the fluid.``

\vspace{3 mm}

\noindent In $\textbf{Eulerian Approach}$ This corresponds to a coordinate system fixed in space, and
in which fluid properties are studied as functions of time as the
flow passes fixed spatial locations.

\vspace{3 mm}

\noindent suppose a fluid property f is function of coordinate (x,y,z,t)
as like f = f(x,y,z,t) and velocity vector $\textbf{U} \equiv (u,v, w)^{T}$ then

\[df = \frac{\partial f}{\partial t}dt + \frac{\partial f}{\partial x}dx + \frac{\partial f}{\partial y}dy +
\frac{\partial f}{\partial z}dz \]

\[\Rightarrow \frac{df}{dt} = \frac{\partial f}{\partial t} + \frac{\partial f}{\partial x}\frac{dx}{dt} + \frac{\partial f}{\partial y}\frac{dy}{dt} +
\frac{\partial f}{\partial z}\frac{dz}{dt}\]

\[\Rightarrow \frac{df}{dt} = \frac{\partial f}{\partial t} + u\frac{\partial f}{\partial x} + v\frac{\partial f}{\partial y} + w\frac{\partial f}{\partial z}\]

\[\Rightarrow \frac{df}{dt} = \frac{\partial f}{\partial t}  + \left(\textbf{U}\cdot\nabla\right) f\]

\noindent the term $\frac{\partial f}{\partial t}$ is the local acceleration and
 $\left(\textbf{U}\cdot\nabla\right)$ the convective acceleration. The convective 
acceleration is nonlinear in f,which is the source of the
great complexity of the mathematics and physics of fluid motion.

\section{General Transport Theorem}

 \normalsize{The general transport
 theorem shows a beautiful equation from which the all conservation equations can be derived. 
 This equation gives a expression for the variation of a physical 
quantity distributed in a scalar or vectorial field inside a deformable volume.
 This deformable volume can have variable mass if its boundary
 is permeable (open system). The basic Reynolds Transport Theorem shows the
 relation for a constant mass volume and a fixed volume, but it can be easily
 expanded to consider changes in a generic volume. The most interesting fact
 about these theorem is that it is only a geometrical relation. Physics only
 gets into the transport theorem when we define changes in a constant mass
 volume, which can use known physical laws, specially from thermodynamics.}

\vspace{5 mm}
\subsection{Control System}
\textbf{Control Mass:}

\begin{enumerate}
 \item It is a system of fixed mass.
\item  This type of system is usually referred to as ``closed system''.
\item There is no mass transfer across the system boundary.
\item Energy transfer may be take place into or out of the system.
\end{enumerate}

\vspace{3 mm}
\textbf{Control Volume:}

\begin{enumerate}
 \item It is a system of fixed Volume.
\item This type of system usually referred to as ``Open System'' or a ''Control Volume``.
\item Mass transfer take place across the control volume.
\item Energy Transfer also occur into or out of the system.

\end{enumerate}

\vspace{5 mm}
\subsection{Reynolds Transport Theorem:}

\vspace{3 mm}

\noindent The question arise as to how a given quantity changes $\psi$ (where, $\psi(\textbf{x},t)$
some fluid property per unit volume) with respect to time as a material volume deforms.

\vspace{2 mm}
\noindent We define the material volume as: An arbitrary chosen control volume of fluid whose surface
moves at the particle velocity.

\vspace{2 mm}
\noindent The significance of the volume moving at the same rate as the particle velocity is that no mass is
transported across the chosen control surface that encloses the control volume. In addition, we
state by definition that the material control volume deforms with the body motion.

\begin{figure}[h]
 \centering
 \includegraphics[width=10cm,height=10cm,bb=0 0 774 431,keepaspectratio=true]{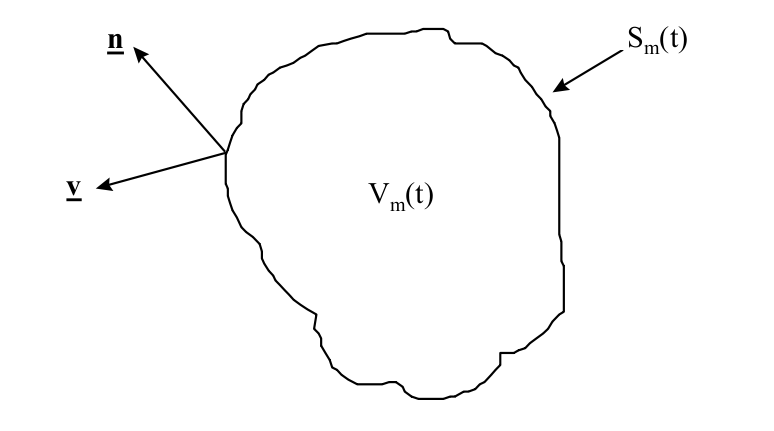}
 \caption{Material Control Volume}
 \label{fig:1}
\end{figure}

\vspace{13 mm}

\noindent Consider a material volume $V_{m}(t)$ with surface area $S_{m}(t)$.
 The unit normal to the surface is denoted by $\underline{\textbf{n}}$  and
 the surface velocity is denoted by $\underline{\textbf{v}}$. But in our calculation
 $\underline{\textbf{n}} \Leftrightarrow \hat{\textbf{n}}$ and
 $\underline{\textbf{v}} \Leftrightarrow \textbf{U}$. Within this control
 volume is a property of the continuum $\psi$ that is of interest. We would
 like to determine how $\psi$ changes with time. The total amount of $\psi$ present
 in the volume $V_{m}(t)$ can be expressed as:
\begin{equation}
 \textbf{F}(t) = \int_{V_{m}(t)}\psi(\textbf{x},t)dV
\end{equation}

\vspace{3 mm}

\noindent The time rate of change if the fluid property is defined as:

\vspace{3 mm}

\begin{equation} \label{eq:2.2}
 \frac{d\textbf{F}}{dt} = \frac{d}{dt}\int_{V_{m}(t)}\psi(\textbf{x},t)dV
\end{equation}

\vspace{2 mm}

\noindent $\textbf{Since integration limit}$  $V_m(t)$ $\textbf{is function of time so we can not take the derivative inside the}$ $\hspace{15mm}$ 
$\textbf{integration directly}$. In order to overcome this 
problem we transform the volume, and material property from a spatial 
representation to a reference/material representation. Recall that any
differential volume element at time t is 

\vspace{2 mm}

\[dV = JdV_{0}\]

\vspace{2 mm}

\noindent where $dV_{0}$ is differential volume at t = 0 and Jacobian J is defined as

\vspace{2 mm}

\begin{equation}\label{eq:2.3}
 J = det F = \begin{bmatrix}
\vspace{2mm}
              \frac{\partial x_{1}}{\partial X_{1}} &
 \frac{\partial x_{2}}{\partial X_{1}} & \frac{\partial x_{3}}{\partial X_{1}} \\

\vspace{2mm}

\frac{\partial x_{1}}{\partial X_{2}} & \frac{\partial x_{2}}{\partial X_{2}}
& \frac{\partial x_{3}}{\partial X_{2}} \\

\frac{\partial x_{1}}{\partial X_{3}} & \frac{\partial x_{2}}{\partial X_{3}}
& \frac{\partial x_{3}}{\partial X_{3}}
             \end{bmatrix}
\end{equation}

\noindent   Given $\textbf{x} = \Phi(\textbf{X},t)$ also $\Psi(\textbf{x},t)$ 
can be expressed in terms of material coordinates by employing the results is

\vspace{2 mm}

\[\Psi(\textbf{x},t) = \Psi({\Phi(\textbf{x},t)},t) = \Psi(\textbf{X},t)\]

\vspace{2 mm}

\noindent Therefore we can express the integral in equation ~\ref{eq:2.2} in terms of the 
material coordinates:

\[\frac{d\textbf{F}}{dt} = \int_{V_{0}}\frac{d}{dt}\left[\Psi(\textbf{X},t)J\right]dV_{0}\]

\[\Rightarrow \frac{d\textbf{F}}{dt} = \int_{V_{0}}\left[\frac{d\Psi(\textbf{X},t)}{dt}J + \Psi(\textbf{X},t)\frac{dJ}{dt}\right]dV_{0}\]

\begin{equation}
 \Rightarrow \frac{d\textbf{F}}{dt} = \int_{V_{0}}
\left[\left(\frac{\partial \Psi(\textbf{X},t)}{\partial t} +
 (\textbf{U}\cdotp\nabla)\Psi\right)J + \Psi(\textbf{X},t)\frac{dJ}{dt}
\right]dV_{0}
\end{equation}

\noindent  Where U is the velocity of control volume  Now we recall J, from
 equation~\ref{eq:2.3} we can say the J is function of
 $\frac{\partial \textbf{x}}{\partial \textbf{X}}$ i.e.

\[J = J\left(\frac{\partial \textbf{x}}{\partial \textbf{X}}\right)\]

\[\Rightarrow dJ = \frac{\partial J}{\partial\left(\frac{\partial\textbf{x}}{\partial \textbf{X}}\right)}d(\frac{\partial \textbf{x}}{\partial \textbf{X}})\]

\[\Rightarrow \frac{dJ}{dt} = \frac{\partial J}{\partial \left(\frac{\partial \textbf{x}}{\partial \textbf{X}}\right)}\frac{d}{dt}\left(\frac{\partial \textbf{x}}{\partial \textbf{X}}\right) \]

\[\Rightarrow  \frac{dJ}{dt} = \frac{\partial J}{\partial\left(\frac{\partial \textbf{x}}{\partial \textbf{X}}\right)}\frac{\partial}{\partial X}\left(\frac{d\textbf{x}}{dt}\right)\]

\begin{equation} \label{eq:2.4}
 \Rightarrow \frac{dJ}{dt} = J\left(\nabla \cdot \textbf{U}\right)
\end{equation}

\noindent   Now putting the value of $\frac{dJ}{dt}$ in equation ~\ref{eq:2.4} we get,

\[ \frac{d\textbf{F}}{dt} = \int_{V_{0}}\left[\frac{\partial \Psi(\textbf{X},t)}{\partial t} +
 (\textbf{U}\cdotp\nabla)\Psi + \Psi(\textbf{X},t)\left(\nabla \cdot \textbf{U}\right)\right]JdV_{0}\]

\[\Rightarrow \frac{d\textbf{F}}{dt} = \int_{V_{m}(t)}\left[\frac{\partial \Psi}{\partial t} + \nabla\cdot\left(\textbf{U}\Psi\right)\right]dV_{m}\]

\begin{equation} \label{eq:2.6}
 \Rightarrow \frac{d}{dt} \int_{V_{m}(t)}\Psi(\textbf{x},t)dV_{m} = \int_{V_{m}(t)}\left[\frac{\partial \Psi}{\partial t} + \nabla\cdot\left(\textbf{U}\Psi\right)\right]dV_{m}
\end{equation}

\vspace{2 mm}

\noindent   This is known as $\textbf{Reynold's Transport Theorem}$.\cite{6} Now applying Gauss's Divergence Theorem in $2^{nd}$ of the integral for the 
surface area $S_{m}(t)$ of the control volume of volume $V_m(t)$ we get

\vspace{2 mm}

\[ \frac{d\textbf{F}}{dt} = \int_{V_{m}(t)}\frac{\partial \Psi}{\partial t}dV_{m} + \int_{S_{m}(t)}\Psi\textbf{U}\cdot \hat{\textbf{n}}dS_{m}\]

\begin{equation} \label{eq:2.7}
 \frac{d}{dt}\int_{V_{m}(t)} \Psi \left(\textbf{x},t\right)dV_{m} = \int_{V_{m}(t)}\frac{\partial \Psi}{\partial t}dV_{m} + \int_{S_{m}(t)}\Psi\textbf{U}\cdot \hat{\textbf{n}}dS_{m}
\end{equation}

$\hspace{63mm}$ Rate of change at$\hspace{10mm}$ Net flow of $\Psi$
\vspace{1 mm}
$\hspace{160mm}$ .$\hspace{68mm}$a point volume $\hspace{10mm}$ across the surface $S_{m}(t)$

\vspace{3 mm}
\noindent   This is another form of $\textbf{RTT}$ which is known as $\textbf{General Transport Theorem}.$

\section{Conservation of Mass}
\noindent   Conservation of mass The constancy of mass is inherent in the definition of a control
 mass system and therefore we can write
\begin{equation}
 \left(\frac{dm}{dt}\right)_{CMS} = 0
\end{equation}

\vspace{3 mm}
\noindent   The total mass of the the control volume can be defined as

\[m = \int_{V_{m(t)}}\rho dV_{m}\]

\vspace{3 mm}
\noindent   To develop the analytical statement for the conservation of mass of 
a control volume, by using equation ~\ref{eq:2.6} we get,

\[\frac{d\textbf{m}}{dt} = \int_{V_{m}(t)}\left[\frac{\partial \rho}{\partial t} + \nabla\cdot\left(\textbf{U}\rho\right)\right]dV_{m}\]
\[\Rightarrow \int_{V_{m}(t)}\left[\frac{\partial \rho}{\partial t} + \nabla\cdot\left(\textbf{U}\rho\right)\right]dV_{m} = 0\]

\begin{equation} \label{eq: 3.2}
 \frac{\partial \rho}{\partial t} + \nabla\cdot\left(\textbf{U}\rho\right) = 0  
\end{equation}

\vspace{3 mm}
\noindent   This known as $\textbf{mass conservation equation}$ also known as 
differential form of continuity equation.

\section{ConservatioFconsern of Momentum}

\noindent Conservation of Momentum or Momentum Theorem The principle of conservation
 of momentum as applied to a control volume is usually referred to as the
 momentum theorem.The first step in deriving the analytical statement of
 linear momentum theorem is to write the equation ~\ref{eq:2.7} for the property $\Psi$ as
 the linear-momentum $m\textbf{u}$  and the velocity $\textbf{u}$ along x-direction .
 Then it becomes

\begin{equation} \label{eq:4.1}
 \frac{d}{dt}\int_{V_{m}(t)}\rho u dV_{m} = \int_{V_{m}(t)}\left[\frac{\partial}{\partial t}\left(\rho u\right) + \nabla\cdot\left(\rho u\textbf{U}\right)\right]dV_{m}
\end{equation}

\[\Rightarrow\frac{d}{dt}\int_{V_{m}(t)}\rho u dV_{m} = \int_{V_{m}(t)}\left[\frac{\partial}{\partial t}\left(\rho u\right) + \nabla\cdot\left(\rho u\textbf{U}\right)\right]dV_{m}\]
\[= \int_{V_{m}(t)} u\left[\frac{\partial \rho}{\partial t} + \nabla\cdot\rho\textbf{U}\right] + \left[\rho\frac{\partial u}{\partial t} + \rho\textbf{U}\cdot\nabla u\right]dV_{m}\]
\vspace{3 mm}
\noindent  The last term of $1^{st}$ term of the integral and first term of $2^{nd}$ term of
the integral create shows the equation of mass conservation which gives zero.
so that

\begin{equation} \label{eq:4.2}
\int_{V_{m}(t)}\left[\rho\frac{\partial \rho}{\partial t} + \rho\textbf{U}\cdot\nabla u\right]dV_{m} = \int_{V_{m}(t)}\rho\frac{du}{dt}dV_{m}
\end{equation}

\vspace{3 mm}
\noindent From equation ~\ref{eq:4.1} and ~\ref{eq:4.2} we get (similar equations can
 be derived for v and w as well)

\vspace{3 mm}

\begin{equation}
 \int_{V_{m}(t)}\rho\frac{d\textbf{U}}{dt}dV_{m} = \int_{V_{m}(t)}\textbf{F}_{B}dV_{m} + \int_{S_{m}(t)}\textbf{F}_{S}dS_{m}
\end{equation}

\vspace{3 mm}

\noindent  We begin by noting that $\textbf{F}_{S}$ be a vector that there must be a 
matrix say $\textbf{T}$ such that

\[\textbf{F}_{s} = \textbf{T}\cdot\hat{\textbf{n}}\]

\vspace{3 mm} 

\noindent  Notice that the ''dot'' notation for the matrix-vector product is used to 
emphasize that each component of $\textbf{F}_{S}$ is the (vector) dot product
 of the corresponding row of $\textbf{T}$ with the unit column vector 
$\hat{\textbf{n}}$. So we get

\vspace{3 mm}

\[\int_{V_{m}(t)}\rho\frac{d\textbf{U}}{dt}dV_{m} = \int_{V_{m}(t)}\textbf{F}_{B}dV_{m} + \int_{S_{m}(t)} \textbf{T}\cdot\hat{\textbf{n}}dS_{m}\]
\vspace{3 mm}
\noindent By using Gauss Divergence Theorem for the last term of RHS and recalling
 $V_{m}(t)$ as arbitrary fluid element, we get

\begin{equation} \label{eq:4.4}
 \rho\frac{d\textbf{U}}{dt} - \textbf{F}_{B} -\nabla\cdot\textbf{T} = 0
\end{equation}
\vspace{3 mm}

\noindent This provides a fundamental and very general momentum balance that is valid at
all points of any fluid flow (following the continuum hypothesis). $\textbf{T}$ is reason of surface force.
Now we are not going to more expression of $\textbf{T}$. Now writing the final form of $\textbf{T}$ 

\begin{equation}
 \textbf{T} = \begin{bmatrix}
\vspace{2 mm}
               -P+\tau_{xx} & \tau_{xy} & \tau_{xz} \\
\vspace{2 mm}
\tau_{yx} & -P+\tau_{yy} & \tau_{yz} \\
\tau_{zx} & \tau_{zy} & -P +\tau_{zz}
              \end{bmatrix}
\end{equation}

\[\Rightarrow \textbf{T} = -P\textbf{I} + \tau\]

\vspace{2 mm}

\noindent In this equation $\textbf{I}$ is the identity matrix, and $\tau$
is often termed the viscous stress tensor. The
expression will be clear from the picture
\begin{figure}[h]
 \centering
 \includegraphics[width=8cm,height=8cm,bb=0 0 584 584,keepaspectratio=true]{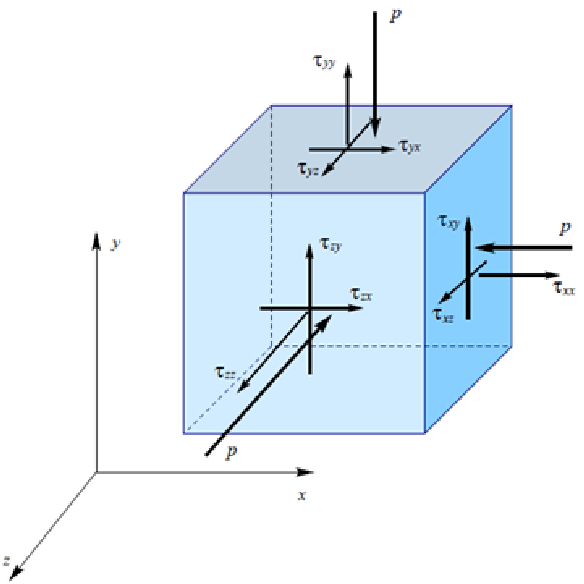}
 \caption{Stress Function}
 \label{fig:2}
\end{figure}

\noindent From Newton's law of viscosity we can express the components of shear stress

\vspace{3 mm}

\[\tau_{xy} = \mu\left(\frac{\partial u}{\partial y} + \frac{\partial v}{\partial x}\right)\]

\vspace{2 mm}
\noindent Putting this value in equation ~\ref{eq:4.4} we get(in component form)

\begin{equation}\label{A_Label}
 \begin{split}
  \rho\frac{du}{dt} = -P_{x} + \mu\left(u_{xx} + u_{yy}+ u_{zz}\right) + F_{B_{,x}} \\
\rho\frac{dv}{dt} = -P_{y} + \mu\left(v_{xx} + v_{yy}+ v_{zz}\right) + F_{B_{,y}} \\
\rho\frac{dv}{dt} = -P_{z} + \mu\left(w_{xx} + w_{yy}+ w_{zz}\right) + F_{B_{,z}}
 \end{split}
\end{equation}

\noindent Now divide these equations by $\rho$ on both side 
 and  express them as
\begin{equation} \label{eq:4.7}
\begin{split}
 \frac{\partial u}{\partial t} + u\frac{\partial u}{\partial x} +v\frac{\partial u}{\partial y} +w\frac{\partial u}{\partial z} = 
\frac{\partial P}{\partial x} + \nu\triangle u + \frac{1}{\rho} + F_{B_{,x}} \\
\frac{\partial v}{\partial t} + u\frac{\partial v}{\partial x} +v\frac{\partial v}{\partial y} +w\frac{\partial v}{\partial z} = 
\frac{\partial P}{\partial y} + \nu\triangle v + \frac{1}{\rho} + F_{B_{,y}} \\
\frac{\partial w}{\partial t} + u\frac{\partial w}{\partial x} +v\frac{\partial w}{\partial y} +w\frac{\partial w}{\partial z} = 
\frac{\partial P}{\partial z} + \nu\triangle w + \frac{1}{\rho} + F_{B_{,z}} 
\end{split}
\end{equation}
\vspace{3 mm}

\noindent Here, $\nu$ is kinematic viscosity, the ratio of viscosity $\mu$ to density
 $\rho$ and $\triangle$ is the second-order partial differential operator
 (given here in Cartesian coordinates)

\vspace{3 mm}
\[\triangle = \frac{\partial^{2}}{\partial x^{2}} + \frac{\partial^{2}}{\partial y^{2}} + \frac{\partial^{2}}{\partial z^{2}}\]
\vspace{3 mm}

\noindent This is known as momentum conservation equation but well known as $\textbf{Nevier-Stokes Equation}$.
and $\textbf{they provide a pointwise
description of essentially any time-dependent incompressible fluid flow.
}$

\section{Energy Conservation}

 \noindent Let E be the total energy of the fluid, sum of its kinetic energy
 and internal energy:
\vspace{3 mm}

\[E = E_{k} + E_{int} = \frac{1}{2}\int_{V_{m}(t)}\rho\textbf{u}^{2}dV_{m} + \int_{V_{m}(t)}\rho\epsilon dV_{m}\]

\vspace{3 mm}

\noindent Principle of energy conservation: “the variation in time of the total energy 
of a portion of fluid is equal to the work done per unit time over the system
 by the stresses (internal forces) and the external forces”. Apply E in $\bf{RTT}$
equation ~\ref{eq:2.6}, we get

\begin{equation} \label{eq:5.1}
 \frac{dE}{dt} = \frac{d}{dt}\int_{V_{m}(t)}\left(\frac{1}{2}\rho\textbf{u}^{2} + \rho\epsilon\right)dV_{m} = \int_{V_{m}(t)}\left[\frac{\partial}{\partial t}\left(\frac{1}{2}\rho \textbf{u}^{2}+\rho\epsilon\right) + \nabla\cdot\left(\frac{1}{2}\textbf{u}\rho \textbf{u}^{2}+\textbf{u}\rho\epsilon\right)\right]dV_{m}
\end{equation}

\vspace{3 mm}

\noindent  Here, $\frac{\partial}{\partial t}\left(\frac{1}{2}\rho u^{2}\right) = \rho u_{i}\frac{\partial u_{i}}{\partial t}$
 and substituting in equation ~\ref{eq:4.7}, we get

\vspace{3 mm}

\[\frac{\partial u_{i}}{\partial t} = -u_{k}\frac{\partial u_{i}}{\partial x_{k}} - \frac{1}{\rho}\frac{\partial P}{\partial x_{i}} + u_{i}\frac{\partial \sigma_{ik}}{\partial x_{k}}\]

\vspace{3 mm}

\noindent  The result is 
\[\frac{\partial}{\partial t}\left(\frac{1}{2}\rho\textbf{u}^{2}\right) = -\rho\textbf{u}\cdot\left(\textbf{u}\cdot\nabla\right)\textbf{u} - \textbf{u}\cdot\nabla P + u_{i}\frac{\partial \sigma_{ik}}{\partial x_{k}}\]
\[\Rightarrow\frac{\partial}{\partial t}\left(\frac{1}{2}\rho \textbf{u}^{2}\right) = -\rho\left(\textbf{u}\cdot\nabla\right)\left(\frac{u^{2}}{2}+\frac{P}{\rho}\right) + \nabla(\textbf{u}\cdot\sigma)-\sigma_{ik}\frac{\partial u_{i}}{\partial x_{k}}\]

\vspace{3 mm}
\noindent Here $\textbf{u}\cdot\sigma$ denotes the vector whose components are 
$u_{i}\sigma_{ik}$. Since $\nabla\cdot\textbf{u} = 0$ for an incompressible
 fluid, we can write the first term on the right as a divergence:

\vspace{3 mm}

\begin{equation}
 \frac{\partial}{\partial t}\left(\frac{1}{2}\rho \textbf{u}^{2}\right) = -\nabla\cdot\left[\rho\textbf{u}\left(\frac{1}{2}\textbf{u}^{2} + \frac{P}{\rho}\right) - \textbf{u}\cdot\sigma\right] - \sigma_{ik}\frac{\partial u_{i}}{\partial x_{k}} 
\end{equation}
\vspace{1 mm}
\hspace{80 mm}$\Downarrow$\hspace{25 mm}$\Downarrow$\hspace{20 mm}$\Downarrow$
\vspace{2 mm}

\hspace{60 mm} $\textbf{Energy Flux}$ \hspace{7 mm}$\textbf{Energy Flux}$ \hspace{5 mm} $\textbf{Energy Flux}$

\vspace{1 mm}

\hspace{60 mm} $\textbf{due to actual}$ \hspace{5 mm} $\textbf{due to Friction}$ \hspace{3 mm} $\textbf{due to momentum}$

\vspace{1 mm}

\hspace{60 mm} $\textbf{mass transfer}$  \hspace{43 mm} $\textbf{change}$

\vspace{3 mm}

\noindent Now from equation ~\ref{eq:5.1}, we get

\begin{equation} \label{eq: 5.3}
 \int_{V_{m}(t)}\left[\frac{\partial}{\partial t}\left(\frac{1}{2}\rho\textbf{u}^{2} + \rho\epsilon\right) + \nabla\cdot\left(\left(\frac{1}{2}
\rho\textbf{u}^{2} + \rho\epsilon + 
P\right)\textbf{u} - \textbf{u}\cdot\sigma\right)\right]dV_{m} = -\int_{V_{m}(t)}\sigma_{ik}\frac{\partial u_{i}}{\partial x_{k}}
\end{equation}
\noindent This is known as $\textbf{Energy conservation equation}$.\cite{2}

\section{Euler Equation:}

\noindent The purpose of this note derivation of Euler's equation for fluid flow on the basis of
conservation of mass, conservation of momentum and Newton's laws.
  
 \noindent To set the stage, consider the example shown in figure 4. The top half is a 
snapshot of the fluid at some initial time $t_{0}$ and the bottom half is a
 snapshot at some slightly later time $t_{1}$. One parcel of fluid has been
 marked with blue dye, and another parcel has been marked with red dye. 
The rectangle represents an imaginary box, also called the ``Control Volume"
we will be particularly interested in what is happening within this box.

\begin{figure}[h]
 \centering
 \includegraphics[width=12cm,height=4cm,bb=0 0 605 280,keepaspectratio=true]{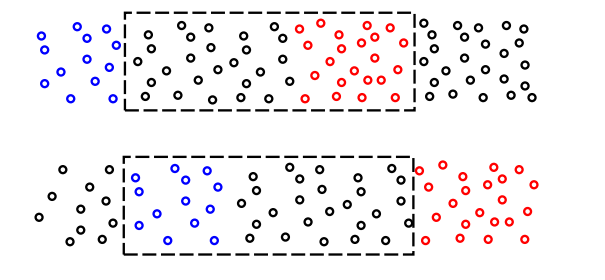}
 \caption{fluid element in imaginary box}
 \label{fig:3}
\end{figure}

\vspace{2 mm}

\noindent There are some qualitative observations that we can make immediately:
\begin{enumerate}
\item[$\ast$] The fluid is flowing left-to-right.
\item[$\ast$] The density depends on position: The red parcel is denser than the blue parcel.
\item[$\ast$]{The density depends on time: The density in the control volume decreases as time
passes from $t_{0}$ to $t_{1}$.}
\item[$\ast$]{The velocity depends on position: during the time interval
 in question, the right edge of the blue parcel moves a relatively short
 distance, while the right edge of the red parcel moves a relatively long
 distance.}
\item[$\ast$]{The total number of red particles does
 not change, and the total number of blue particles does not change. (Some
 particles that flow out of the frame of the diagram are not accounted for,
 but all particles  -red, blue, or black-that affect the control volume are
 accounted for.}
\end{enumerate}

\vspace{2 mm}

\noindent In our calculations, we consider the following assumptions 
\begin{enumerate}
 \item[\ddag] The fluid is ideal gas.
\item[\ddag] There is no viscous force i.e $\mu=0$
\item[\ddag] There is no external force.
\item[\ddag] Fluid flow is one dimensional.
\end{enumerate}

\vspace{2 mm}

\noindent Now applying this assumption on equation ~\ref{eq: 3.2}

\vspace{2 mm}

\[\frac{\partial \rho}{\partial t} + \nabla\cdot\left(\rho\textbf{u}\right) = 0\]
\begin{equation}
 \Rightarrow\frac{\partial \rho}{\partial t} + \frac{\partial\left(\rho u\right)}{\partial x} = 0 
\end{equation}

\noindent This is $\textbf{Mass Conservation}$ equation for ideal gas.
\vspace{3 mm}

\noindent $\star$ Momentum Conservation: In some volume of fluid with surface
 $S_{m}(t)$ the total force acting on the volume $V_{m}(t)$ is 
$-\int_{S_{m}(t)}PdS_{m} = -\int_{V_{m}(t)}\nabla P dV_{m}$ where P is the pressure.
Applying force on unit volume $\rho\frac{d\textbf{u}}{dt}$ (Newton's $2^{nd}$ Law).
Then we  may be written as 

\vspace{2 mm}
\[\rho\frac{d\textbf{u}}{dt} = -\nabla P\]
\vspace{2 mm}
By using $\textbf{RTT}$ equation ~\ref{eq:2.6}, we get
\[\frac{\partial(\rho\textbf{u})}{\partial t} + \nabla\cdot\left[\textbf{u}\otimes\left(\rho\textbf{u}\right)\right] = 0\]

\[\Rightarrow \frac{\partial \left(\rho u_{j}\right)}{\partial t} + \sum_{i=1}^{3}\frac{\partial \left(\rho u_{i}u_{j}\right)}{\partial x_{i}} + \frac{\partial P}{\partial x_{j}} = 0\]
But for one dimensional case $i=j=1$, then
\begin{equation}
 \frac{\partial\left(\rho u\right)}{\partial t} + \frac{\partial\left(\rho u^{2} + P\right)}{\partial x} = 0 
\end{equation}

\noindent This is $\textbf{Momentum conservation}$  
\vspace{3 mm}
and now applying the assumption on energy equation ~\ref{eq: 5.3}:

\[\int_{V_{m}(t)}\left[\frac{\partial\left(\rho \textbf{u}^{2} + \rho\epsilon\right)}{\partial t} + \nabla\cdot\left(\left(\frac{1}{2}\rho\textbf{u}^{2} + \rho\epsilon + P\right)\textbf{u}\right)\right]dV_{m} = 0\]
\vspace{3 mm}
recalling arbitrary value of elementary volume and applying one dimensional case

\vspace{2 mm}

\[\Rightarrow \frac{\partial\left(\rho u^{2} + \rho\epsilon\right)}{\partial t} + \frac{\partial }{\partial x}\left[\left(\frac{1}{2}\rho u^{2} + \rho\epsilon +P\right)u\right] = 0\]
\vspace{2 mm}
 Total energy of ideal gas $e = \frac{1}{2}\rho u^{2} + \rho\epsilon$, so
\begin{equation}
 \frac{\partial e}{\partial t} + \frac{\partial}{\partial x}\left[u\left(e + P\right)\right] = 0
\end{equation}

\noindent This is known as $\textbf{Energy Conservation}$ equation for ideal gas.
\vspace{2 mm}

\section{Finite Volume Method}

\noindent The finite-volume method\cite{5} is a method for representing and evaluating partial differential
equations in the form of algebraic equations. In this method values are calculated from at discrete places 
at meshed geometry. "Finite Volume" refers small volume surroundings about mesh grid point. The evaluated terms are 
Fluxes at the surface of referred volume. Because the flux entering a given volume is identical to that leaving
the adjacent volume, these methods are conservative. This method is used in many $\textbf{CFD}$ packages.

\noindent \normalsize{Let us consider a 1 Dimensional Hyperbolic Equation can be defined by partial differential equation }

\begin{equation}
 \frac{\partial\rho}{\partial t} + \frac{\partial \Psi}{\partial x} = 0 , \hspace{7mm} t \geq 0 
\end{equation}

\noindent  Where, $ \rho = \rho(x,t) $ represents the state variable  and  $\Psi = \Psi(x,t)$ represents the flux
or flow of $\rho$ .Conventionally, positive $\Psi$ indicates flow from left to right and negative $\Psi$ indicates
flow from right to left.If we assume that equation (7.1) represents a flowing medium of constant area, we can sub-divide
the spatial domain x, into finite volumes or cells with cell center index i. For a particular cell 'i' we can define
the volume average value of $\rho_{i}(t) = \rho(x,t)$ at time t = $t_{1}$ and $x \in [x_{i-1/2}, x_{i+1/2}]$. Just look at the 
figure

\begin{figure}[h]
 \centering
 \includegraphics[width=12cm,height=10cm,bb=0 0 665 101,keepaspectratio=true]{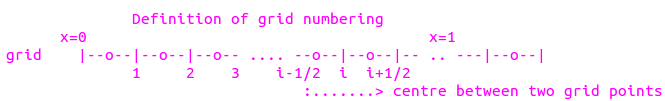}
 \caption{A simple one dimensional grid}
 \label{fig:4}
\end{figure}

\noindent Now average value equation is 

\begin{equation}
\bar{\rho_{i}}(t_{1})  =  \frac{1}{x_{i+1/2} - x_{i-1/2}} \int_{x_{i-1/2}}^{x_{i+1/2}}\rho(x,t_{1})dx
\end{equation}

\noindent At time $t = t_{2}$, the average value equation is

\begin{equation}
\bar{\rho_{i}}(t_{2})  =  \frac{1}{x_{i+1/2} - x_{i-1/2}} \int_{x_{i-1/2}}^{x_{i+1/2}}\rho(x,t_{2})dx
\end{equation}

\noindent Where, $x_{i-1/2}$ and $x_{i+1/2}$ represent locations of the upstream and downstream faces or edges respectively
of the $i^{th}$ cell. Now integrating equation(7.1) with respect to time, we have:

\begin{equation}
 \rho_{i}(t_{2})  = \rho_{i}(t_{1}) -  \int_{t_{1}}^{t_{2}}\Psi_{x}(x,t)dt
\end{equation}

\noindent where $\Psi_{x} = \frac{\partial\Psi}{\partial x}$, To obtain the volume average  $\rho(x,t)$ at time
$t = t_{2}$ we integrate $\rho(x,t_{2})$ over the cell volume $[x_{i-1/2},x_{i+1/2}]$ after then we divide the result by
 $\triangle x_{i} = x_{i+1/2} - x_{i-1/2}$ , implies that
 
\vspace{2 mm}
\begin{equation}
 \bar{\rho_{i}}(t_{2}) =  \frac{1}{\triangle x_{i}}\int_{x_{i-1/2}}^{x_{i+1/2}} \textbf{[} \rho(x,t_{1}) - \int_{t_{1}}^{t_{2}}\Psi_{x}(t,x) \textbf{]}dx
\end{equation}

\vspace{2 mm}

\noindent{we can assume that $\Psi$ is well behaved and flow is normal to the surface area, even for 1 D case
$\Psi_{x} \sim \bigtriangledown \Psi$, so we can apply gauss divergence theorem in equation(7.5) over the x region
$[x_{i-1/2}, x_{i+1/2}]$ even x and t are independent to each other i.e. we can interchange the integration of
dx and dt, now we have:}

\begin{equation}
 \bar{\rho_{i}}(t_{2}) = \frac{1}{\triangle x_{i}}\int_{x_{i-1/2}}^{x_{i+1/2}}\rho(x,t_{1})dx - 
 \frac{1}{\triangle x_{i}}\int_{t_{1}}^{t_{2}}\textbf{[}\int_{x_{i-1/2}}^{x_{i+1/2}}\Psi_{x}(t,x)dx\textbf{]}dt
\end{equation}

\begin{equation} \label{eq:7.7}
 \bar{\rho_{i}}(t_{2}) = \bar{\rho}(x,t_{1}) - 
 \frac{1}{\triangle x_{i}}\int_{t_{1}}^{t_{2}}\textbf{[}\Psi_{x_{i+1/2}} - \Psi_{x_{i-1/2}}\textbf{]}dt
\end{equation}

\noindent We can therefore derive semi discrete numerical method with cell center index i and edges at $i \pm 1/2$ :

\begin{equation}
 \frac{d\bar{\rho_i}}{dt} = \frac{1}{\triangle x_{i}} {\textbf{[}{\Psi_{i+1/2} - \Psi_{i-1/2}}\textbf{]}}
\end{equation}

\noindent where values for the edge fluxes, $\Psi_{i+1/2}, \Psi_{i-1/2}$ can be reconstructed by interpolation
or extrapolation of the cell averages. Equation (7.8) is exact for the volume averages; i.e., no approximations have
been made during its derivation.

\section{Roe Scheme Method}

\normalsize{Roe Scheme Method\cite{1} is a method to solve fluid conservation laws which are based on exploiting
the information obtained by considering a sequence of Riemann Problem. It is argued that in existing schemes
much of this information is degraded, and that only certain features of the exact solution are worth striving.}

\noindent we consider the initial value problem for a hyperbolic system of conservation laws, such that

\begin{equation} \label{eq: 8.1}
 \textbf{u}_{t} + \textbf{F}_{x}  = 0
\end{equation}

\noindent where, $\textbf{u} = \textbf{u}(x,t)$ and

\begin{equation} \label{eq: 8.2}
 \textbf{u}(x,0) = \textbf{u}_{0}(x)
\end{equation}

\noindent Where, $\textbf{F}$ is some vector-valued function of $\textbf{u}$, such that the Jacobian matrix 
$A = \frac {\partial \textbf{F}}{\partial \textbf{u}}$ has only real eigenvalues, by using the discrete representation 
$x_{i} = x_{0} + i\triangle x$ and $t_{n} = t_{0} + n\triangle t$ and consider $\textbf{u}_{i}^{n}$ is some approximation of 
$\textbf{u}(x_{i},t_{n})$ but a multitude of strategies are required to obtain the numerical result which is unclear, so we recall
Riemann Problem and condition are:

\begin{equation} \label{eq: 8.3}
 \textbf{u}(x,0) \equiv \textbf{u}_{L} \hspace{7mm} (x < 0);\hspace{10mm} \textbf{u}(x,0) = \textbf{u}_{R} \hspace{7mm}     (x > 0)
\end{equation}

\noindent Now we can solve exactly equation ~\ref{eq: 8.1} by using this initial condition and corresponding system equation
where equation(8.1) will be converted to $\textbf{u}_{t} + A\textbf{u}_{x}  = 0$, which is our exact solution, but now we try to 
solve the equation(8.1) by boundary value and which gives us the approximate solution.

\noindent we consider the approximate solutions

\begin{equation} \label{eq: 8.4}
  \textbf{u}_{t} + \tilde{A}\textbf{u}_{x}  = 0
\end{equation}

\noindent where, $\tilde{A} = \tilde{A}(\textbf{u}_{L}, \textbf{u}_{R})$ which satisfy the properties:

\begin{enumerate}

 \item It constitutes a linear mapping from the vector space $\textbf{u}$ to the vector space $\textbf{F}$

 \item As $\textbf{u}_{L} \textbf{u}_{R} \rightarrow \textbf{u}$, $\bar{A}(\textbf{u}_{l}, \textbf{u}_{R})
 \rightarrow A(\textbf{u})$, where $A = \frac{\partial \textbf{F}}{\partial \textbf{u}}$

 \item For any $\textbf{u}_{L}$,$\textbf{u}_{R}$, $\tilde{A}(\textbf{u}_{L},\textbf{u}_{R}) \times (\textbf{u}_{L} - 
 \textbf{u}_{R}) = \textbf{F}_{L} - \textbf{F}_{R}$

 \item The eigenvectors of $\tilde{A}$ are linearly independent.
\end{enumerate}

\subsection{Parameter Vector:}

\noindent In this section we will use experience in analytic geometry that a plane curve y(x) may in some 
cases be much more easily described by a parametric form y = y(t) and x = x(t). we may therefore expect that a multidimensional 
manifold such as $\textbf{F}(\textbf{u})$ may sometimes be more amenable if represented as $\textbf{F} = \textbf{F}(\textbf{w})$,
$\textbf{u} = \textbf{u}(\textbf{w})$ where $\textbf{w}$ as a parameter vector. Now we exhibit parameter vector for the 
$\textbf{Euler Equations}$, which we write as

\begin{equation} \label{eq: 8.5}
 \textbf{u}_{t} + \textbf{F}_{x} = 0,
\end{equation}

\noindent where

\begin{equation} \label{eq: 8.6}
\boxed{
 \textbf{u} = \begin{bmatrix}
               \rho \\
               \rho u \\
               e \\
               
              \end{bmatrix}
, \hspace{15mm}              \textbf{F} = \begin{bmatrix}
                            \rho u \\
                            P + \rho u^{2} \\
                            u(P + e) \\
                           \end{bmatrix}}
\end{equation}

\noindent in which $\rho$ = density, P = static pressure, u = velocity in Cartesian coordinate x, and e is total energy 
per unit volume. For ideal gas equation :

\begin{equation} \label{eq: 8.7}
 P = \rho RT = \rho(C_{P} - C_{V})T  = \rho(\gamma - 1)C_{V}T = \rho(\gamma - 1)\bar{e}
\end{equation}

\noindent $\bar{e}$ = internal energy per unit mass, so internal energy per unit volume 
$\rho \bar{}\bar{e} = P/(\gamma - 1)$so Total energy = Kinetic energy + Thermal energy i.e.

\begin{equation} \label{eq: 8.8}
 \boxed{e = \frac{1}{2} \rho u^{2} + P/(\gamma -1)}
\end{equation}

\noindent Now we take the parameter vector as like $\textbf{w} = \rho^{\frac{1}{2}}(1, u, h)^{T}$ where h = enthalpy is defined as

\begin{equation} \label{eq: 8.9}
\rho h = e + P = \frac{1}{2}\rho u^{2} +  \frac{P}{(\gamma -1)} + P
\end{equation}

\begin{equation} \label{eq: 8.10}
 \boxed{h - \frac{u^{2}}{2} = \frac{\gamma P}{(\gamma-1)\rho}}
\end{equation}

\begin{equation} \label{eq: 8.11}
 \boxed{(\gamma -1)(h - \frac{u^{2}}{2}) = \frac{\gamma P}{\rho}}
\end{equation}

\noindent Now from the expression of $\textbf{w}$ and from equation ~\ref{eq: 8.6}, we have

\begin{equation} \label{eq: 8.12}
 u_{1} = w_{1}^{2} , \hspace{7mm} u_{2} = w_{1}w_{2} ,\hspace{7mm} u_{3} = \frac{w_{1}w_{3}}{\gamma} + \frac{\gamma -1}{2 \gamma} w_{2}^{2}
\end{equation}

\noindent and

\[F_{1} = w_{1}w_{2} \]

\[F_{2} = P + \rho u^{2}\] 

\[ = \frac{\gamma -1}{\gamma}(\rho h - \frac{\rho u^{2}}{2}) + \rho u^{2}\]

\[ = \frac{\gamma -1}{\gamma} \rho h + \frac{\gamma +1}{2\gamma} \rho u^{2}\]

\[ = \frac{\gamma -1}{\gamma} w_{1}w_{3} + \frac{\gamma + 1}{2\gamma}w_{2}^{2}\]

\begin{equation} \label{eq: 8.13}
 F_{3} = u(P + e) = u \rho h = w_{2}w_{3}
\end{equation}

\subsection{Eigenvalues and Eigenvectors for Euler Equation}

\noindent It is now very easy to represent any jump in space $\textbf{u}$,$\textbf{F}$ in terms of its image in the space $\textbf{w}$. For example,
given any pair of states $(\textbf{u}_{L}, \textbf{u}_{R})$ and their images $(\textbf{w}_{L}, \textbf{w}_{r})$ we can write

\begin{equation} \label{eq: 8.14}
 \left(\textbf{u}_{L} - \textbf{u}_{R}\right)  \equiv \tilde{\textbf{B}}\left(\textbf{w}_{L} - \textbf{w}_{R}\right)
\end{equation}

\noindent and

\begin{equation} \label{eq: 8.15}
 \left(\textbf{F}_{L} - \textbf{F}_{R}\right) \equiv \tilde{\textbf{C}}\left(\textbf{w}_{L} - \textbf{w}_{R}\right)
\end{equation}

\noindent where $\tilde{\textbf{B}}$,$\tilde{\textbf{C}}$ are $3\times3$ matrix, now $\tilde{\textbf{B}}$ can be evaluated by the process:

\[u_{L1} - u_{R1}) = \tilde{B}_{11}\left(w_{L1}-w_{R1}\right) + \tilde{B}_{12}\left(w_{L2} - w_{R2}\right) + \tilde{B}_{13}\left(w_{L3} - w_{R3}\right)\]

\[(w_{L1}^{2} - u_{R1}^{2}) = \tilde{B}_{11}(w_{L1}-w_{R1}) + \tilde{B}_{12}(w_{L2} - w_{R2}) + \tilde{B}_{13}(w_{L3} - w_{R3})\]

\vspace{3 mm}

\noindent So there is only one solution for $\tilde{B}_{11} = \frac{w_{L1}^2 - w_{L2}^{2}}{w_{L1} - w_{R1}} = \bar{w}_{1}$ where 
$\bar{w}_{1} = \frac{1}{2}(w_{L1} + w_{R1})$ and $\tilde{B}_{12}$,$\tilde{B}_{13}$ are zero. Similarly we evaluate another component of
$\tilde{\textbf{B}}$

\[u_{L2} - u_{R2} = \tilde{B}_{21}(w_{L1}-w_{R1}) + \tilde{B}_{22}(w_{L2} - w_{R2}) + \tilde{B}_{23}(w_{L3} - w_{R3})\]

\[\Rightarrow w_{L1}w_{L2} - w_{R1}w_{R2} = \tilde{B}_{21}(w_{L1}-w_{R1}) + \tilde{B}_{22}(w_{L2} - w_{R2}) + \tilde{B}_{23}(w_{L3} - w_{R3})\]

\vspace{3 mm}

\noindent Here $\tilde{B}_{23} = 0$, because LHS have no term of $w_{L3}$ or $w_{R3}$. Now we try to solve the reduced equation. There are two
solution but we have to choose the solution which containing $w_{L}$ and $w_{R}$ both. So the reduced equation is

\[w_{L1}w_{L2} - w_{R1}w_{R2} = \tilde{B}_{21}(w_{L1}-w_{R1}) + \tilde{B}_{22}(w_{L2} - w_{R2})\]

\noindent Where $\tilde{B}_{21} = \frac{1}{2}(w_{L2} + w_{R2})  = \bar{w}_{2}$ and $\tilde{B}_{22} =$ 
$\frac{1}{2}(w_{L1} + w_{R2}) = \bar{w}_{1}$, similarly solving the third equation we have the matrix $\tilde{\textbf{B}}$ as

\begin{equation} \label{eq: 8.16}
 \tilde{\textbf{B}} = \begin{bmatrix}\vspace{3mm}
2 \bar{w}_{1} & 0 & 0 \\
\vspace{3mm}\bar{w}_{2} & \bar{w}_{1} & 0 \\
\frac{\bar{w_{3}}}{\gamma} & \frac{\gamma -1}{\gamma} \bar{w_{2}} & \frac{\bar{w}_{1}}{\gamma} \\
\end{bmatrix}
\end{equation}

\noindent and by using equivalent method

\begin{equation} \label{eq: 8.17}
 \tilde{\textbf{C}} = \begin{bmatrix}
                       \vspace{3mm}\bar{w}_{2} & \bar{w}_{1} & 0 \\
\vspace{3mm}\frac{\gamma -1}{\gamma}\bar{w}_{3} & \frac{\gamma +1}{\gamma}\bar{w}_{2} & \frac{\gamma -1}{\gamma}\bar{w}_{1} \\
0 & \bar{w}_{3} & \bar{w}_{2}
                      \end{bmatrix}
\end{equation}

\noindent Obviously these very simple formulas are closely related to the homogeneous property of the Euler equations. We will wish to
 construct the eigenvalues and the eigenvectors of some matrix $\tilde{A}$ which maps $\triangle \textbf{u}$ and $\triangle \textbf{F}$ with
property $\textbf{u}$. Evidently we may choose $\tilde{A} = (\tilde{\textbf{C}})(\tilde{\textbf{B}})^{-1}$. To find the find the eigenvalues and 
eigenvectors we simplify the matrix $\tilde{\textbf{B}}$ and $\tilde{\textbf{C}}$ divide by $\bar{w}_{1}$ without loss of property.

\vspace{3 mm}
\begin{equation} \label{eq: 8.18}
\boxed{\frac{\bar{w}_{2}}{\bar{w}_{1}} = \frac{\rho_{L}^{\frac{1}{2}}u_{L} + \rho_{R}^{\frac{1}{2}}u_{R}}{\rho_{L}^{\frac{1}{2}} + 
\rho_{R}^{\frac{1}{2}}} =  \bar{u}} 
\end{equation}

\vspace{2 mm}
\begin{equation} \label{eq: 8.19}
 \boxed{\frac{\bar{w}_{3}}{\bar{w}_{1}} = \frac{\rho_{L}^{\frac{1}{2}}h_{L} + \rho_{R}^{\frac{1}{2}}h_{R}}{\rho_{L}^{\frac{1}{2}} + 
\rho_{R}^{\frac{1}{2}}} =  \bar{h} }
\end{equation}

\noindent we use $\textbf{Python}$ code to determine the eigenvalues and eigenvector of new $\tilde{\textbf{B}}$ and $\tilde{\textbf{C}}$:

\vspace{15 mm}
\lstset{
language=Python,
basicstyle=\ttfamily,
otherkeywords={self},             
keywordstyle=\ttfamily\color{blue!90!black},
keywords=[2]{True,False},
keywords=[3]{ttk},
keywordstyle={[2]\ttfamily\color{blue!80!orange}},
keywordstyle={[3]\ttfamily\color{red!80!orange}},
emph={MyClass,__init__},          
emphstyle=\ttfamily\color{red!80!black},    
stringstyle=\color{green!80!black},
showstringspaces=false            
}

\begin{lstlisting}
#!/usr/bin/env python
# -*- coding: utf-8 -*-

from sympy import S, pprint,symbols, simplify
from sympy.matrices import *
from sympy.interactive.printing import init_printing
init_printing = init_printing(use_unicode=False,
  wrap_line=False,no_global=True) 

u, h, gamma = symbols('u h gamma')
gammaA = ((gamma-1)/gamma)
gammaB = ((gamma+1)/gamma)
C = Matrix([[u, 1, 0], [gammaA*h, gammaB*u, gammaA], [0, h, u]])
BP = Matrix([[2, 0, 0], [u, 1, 0], [h/gamma, gammaA*u, 1/gamma]])
B = Inverse(BP)
A = C*B
lam = A.eigenvals()
vec = A.eigenvects()
\end{lstlisting}

\vspace{15 mm}

\noindent Corresponding results are :

\begin{equation} \label{eq: 8.20}
 \tilde{A} = \begin{bmatrix}
              \vspace{3mm}0 & 1 & 0 \\
\vspace{3mm}\frac{\bar{u}^{2}}{2\gamma}(\gamma^{2}- 3\gamma) & \frac{-\bar{u}(\gamma^{2}-3\gamma)}{\gamma} & \gamma -1 \\
\bar{u}(\bar{a}^{2}-\gamma \bar{h}) & -2\bar{a}^{2} -\bar{h} + 2\gamma \bar{h} & \gamma \bar{u}
             \end{bmatrix}
\end{equation}

\vspace{5 mm}
\begin{equation} \label{eq: 8.21}
 \boxed{\lambda_{1} = \bar{u},\hspace{10mm} \lambda_{2} = \bar{a}, \hspace{10mm} \lambda_{3} = \bar{u} + \bar{a}}
\end{equation}

\begin{equation} \label{eq: 8.22}
\boxed{
 e_{1} = \begin{bmatrix}\vspace{3 mm}
           1 \\
\vspace{3 mm}\bar{u} - \bar{a} \\
\bar{h} - \bar{u}\bar{a} \\
          \end{bmatrix}
, \hspace{5 mm} e_{2} = \begin{bmatrix}\vspace{3 mm}
                        1 \\
\vspace{3 mm}\bar{u} \\
\frac{\bar{u}^{2}}{2}
                       \end{bmatrix}
, \hspace{5 mm} e_3 = \begin{bmatrix}\vspace{3 mm}
                      1 \\
\vspace{3 mm}\bar{u} + \bar{a} \\
\bar{h} + \bar{u}\bar{a}\\
                     \end{bmatrix}}
\end{equation}

\vspace{3 mm}

\noindent Here $ \bar{a}^{2} = (\gamma -1)(\bar{h} - \bar{u}^{2}/2) = \frac{\gamma \bar{P}}{\bar{\rho}}$ (by using equation ~\ref{eq: 8.11} ), $\lambda$ 's are eigenvalues and $e_{i}$ are
eigenvectors. 

\subsection{Approximate Riemann Solver}

\noindent  To complete the analysis we must show how to project an arbitrary $\triangle u$ onto the eigenvectors as basis i.e., we have to find
coefficient $\alpha_{i}$ in

\begin{equation} \label{eq: 8.23}
 \triangle u =  \sum_{i=1}^{3} \alpha_{i}e_{i}
\end{equation}

\vspace{3 mm}

\noindent A routine calculation yields
 
\[\triangle u_{1} = \alpha_{1} + \alpha_{2} + \alpha{3}\]

\vspace{2 mm}

\[\triangle u_{2} = \alpha_{1}(\bar{u} -\bar{a}) + \alpha_{2} \bar{u} + \alpha_{3}(\bar{u} +\bar{a})\]

\[\Rightarrow \triangle u_{2} = (\alpha_{1} + \alpha_{2} + \alpha{3})\bar{u} + (\alpha_{3} - \alpha_{1})\bar{a} \]

\[\Rightarrow \triangle u_{2} - \bar{u}\triangle u_{1} = (\alpha_{3} - \alpha_{1})\bar{a} \]

\vspace{2 mm}

\[\triangle u_{3} = \alpha_{1}(\bar{h} - \bar{u}\bar{a}) + \alpha_{2}\frac{\bar{u}^{2}}{2} + \alpha_{3}(\bar{h} + \bar{u}\bar{a})\]

\[\Rightarrow \triangle u_{3} = (\alpha_{3} - \alpha_{1})\bar{u}\bar{a} + \alpha_{2}\frac{\bar{u}^{2}}{2} + 
(\alpha_{1} + \alpha_{3})\bar{h}\]

\[\Rightarrow \triangle u_{3} = (\triangle u_{2} - \bar{u}\triangle u_{1})\bar{u} + \alpha_{2}\frac{\bar{u}^{2}}{2} +
 (\triangle u_{1} -\alpha_{2})\bar{h}\]

\[\Rightarrow \triangle u_{3} = (\bar{h} -\bar{u}^{2})\triangle u_{1} + \alpha_{2}\frac{u^{2}}{2} + \bar{u}\triangle u_{2}\]

\[\Rightarrow \frac{\bar{a}^{2}}{\gamma -1}\alpha{2} = (\bar{h} - \bar{u}^{2})\triangle u_{1} + u\triangle u_{2} - \triangle u_{3}\]

\[\Rightarrow \frac{a^{2}}{\gamma -1}\alpha{2} = (\frac{\bar{a}^{2}}{\gamma -1} -\frac{\bar{u}^{2}}{2})\triangle u_{1} +
 \bar{u}\triangle u_{2} - \triangle u_{3} \]

\begin{equation} \label{eq: 8.24}
\Rightarrow  \boxed{\alpha_{2} = (1 - \frac{(\gamma-1)\bar{u}^{2}}{2\bar{a}^{2}})\triangle u_{1} + \frac{(\gamma - 1)
\bar{u}}{\bar{a}^{2}}\triangle u_{2} - \frac{\gamma -1}{\bar{a}^{2}}\triangle u_{3}}
\end{equation}

\vspace{5 mm}

\noindent  Now from the $1^{st}$ equation :

\[\alpha_{1} + \alpha_{3} = \triangle u_{1} - \alpha_{2}\]
\[\Rightarrow \alpha_{1} + \alpha_{3} = \frac{(\gamma-1)\bar{u}^{2}}{2\bar{a}^{2}}\triangle u_{1} - 
\frac{(\gamma-1)\bar{u}}{\bar{a}^{2}}\triangle u_{2} + \frac{(\gamma -1)}{\bar{a}^{2}}\triangle u_{3}\]

\vspace{3 mm}

\noindent and from $2^{nd}$ equation:

\[-\alpha_{1} + \alpha_{3} = \frac{1}{a}(\triangle u_{2} - u\triangle u_{1})\]
\vspace{3 mm}

\noindent So we are able to solve $\alpha_{1}$ and $\alpha_{3}$:

\vspace{2 mm}
\begin{equation} \label{eq: 8.25}
 \boxed{\alpha_{1} = \frac{\bar{u}}{4\bar{a}}\left[2 + \frac{(\gamma - 1)\bar{u}}{\bar{a}}\right]\triangle u_{1} - \frac{1}{2\bar{a}}
 \left[1 + \frac{(\gamma-1)\bar{u}}{\bar{a}}\right]\triangle u_{2} + \frac{\gamma -1}{2\bar{a}^{2}} \triangle u_{3}}
\end{equation}

\begin{equation} \label{eq: 8.26}
 \boxed{\alpha_{3} = -\frac{\bar{u}}{4\bar{a}}\left[2 + \frac{(\gamma - 1)\bar{u}}{\bar{a}}\right]\triangle u_{1} + \frac{1}{2\bar{a}}
 \left[1 - \frac{(\gamma-1)\bar{u}}{\bar{a}}\right]\triangle u_{2} + \frac{\gamma -1}{2\bar{a}^{2}} \triangle u_{3}}
\end{equation}

\vspace{2 mm}
\noindent Now we construct $\triangle F$, which will construct by the eigenvalues and eigenvectors of $\tilde{A}$ which maps $\triangle u$ and
$\triangle F$ and must be multiplied by $\alpha_{i}$ 's because $\tilde{A}$ in image space $\textbf{w}$. So we have:

\vspace{1 mm}

\[\triangle F = \sum_{i}\lambda_{i}\alpha_{i}e_{i}\]

\vspace{1 mm}

\noindent Now we have have the information $F_{i}$ and $F_{i+1}$ on the mesh grid point so intermediate Flux is given by:

\vspace{2 mm}

\[F_{i+1/2} = \frac{1}{2}(F_{i+1} + F_{i}) - \frac{1}{2}\sum{\lambda_{i}\alpha_{i}e_{i}}\]

\begin{center}
\begin{equation} \label{eq: 8.27}
\Rightarrow F_{i+1/2} = \frac{1}{2}(F_{L} + F_{R}) - \frac{1}{2}\sum{\lambda_{i}\alpha_{i}e_{i}}
\end{equation}
\end{center}

\vspace{2 mm}

\subsection{Harten Sonic Entropy Fix}

\noindent when,  $\bar u = \bar a$  then  $\lambda_{1} = \bar u - \bar a = 0 $  so

\vspace{2 mm}

\[\Rightarrow F_{i+1/2} = \frac{1}{2}(F_{L} + F_{R})\]

\vspace{2 mm}

\noindent This means that central discretization is used, without artificial viscosity so that there is a
danger that entropy decreases, and that discontinuities are insufficiently  smear out. The flow 
is not sonic , then we do not have central discretization , and there may be enough dissipation
 to prevent violation of the entropy condition. this is confirmed by the correct approximation 
by the subsonic fans.

\noindent An often used artifice to add dissipation to the Roe scheme near sonic condition has been 
proposed by Harten(1984). The eigenvalues $\lambda_{1}$ and $\lambda{3}$ are slightly increased in the
vicinity of zero, replacing them in value $\lambda_{i}$'s by

\[\boxed{\mid \tilde{\lambda}_{i}\mid = \mid \lambda_{i}\mid  \hspace{5mm} if \hspace{3mm}  \mid \lambda_{i}\mid  \geq \epsilon, \hspace{3mm} i = 1,3 ;}\]

\begin{equation} \label{eq: 8.28}
 \boxed{\mid \tilde{\lambda}_{i}\mid = \frac{1}{2}(\frac{\lambda_{i}^{2}}{\epsilon} + \epsilon) \hspace{5mm} if \hspace{3mm}  \mid \lambda_{i}\mid  < \epsilon, \hspace{3mm} p = i,3 ;}
\end{equation}

\vspace{2 mm}

\noindent for some small value of $\epsilon$. Since equation ~\ref{eq: 8.21} becomes into play when $\bar  u \approx \bar a$
this is called sonic entropy fix.\cite{8} For super sonic tube $\epsilon = 0.5$ 
corresponding numerical solution is given in section \ref{sec:Appendix}.

\newpage
\section{One-Dimensional Shock-Tube Problem}

\noindent shock tube is a very long pipe in which a strong shock wave is generated.
In the initial configuration, the tube is divided by a diaphragm into two compartments:
the driver section containing the gas with the higher pressure $P_{1}$ and the driven section
with the gas at the lower pressure $P_{4}$.Here, we  omitting the parts about reflected shock
and expansion waves as well as some theory considered too general.
The right-running normal shock is treated on page 21 and
the solution to the complete problem is given on
page 21.

\begin{figure}[h]
 \centering
 \includegraphics[width=12cm,height=10cm,bb=0 0 885 553,keepaspectratio=true]{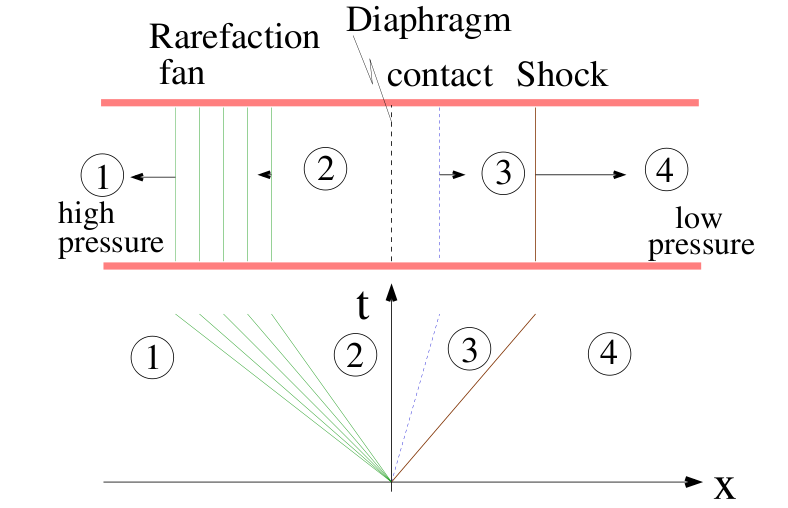}
 \caption{Shock Tube after the Diaphragm Broken} \label{fig: 6}
 \label{fig:5}
\end{figure}

\noindent We start from the continuity, momentum and energy equations
for a stationary normal shock wave\cite{3}:

\begin{equation} \label{eq:9.1}
 \rho_{4}u_{4} = \rho_{3}u_{3}
\end{equation}

\begin{equation} \label{eq:9.2}
 P_{4} + \rho_{4}u_{4}^{2} = P_{3} + \rho_{3}u_{3}^{2}
\end{equation}

\begin{equation} \label{eq:9.3}
 h_{4} +\frac{u_{4}^{2}}{2} = h_{3} + \frac{u_{3}^{2}}{2}
\end{equation}

\noindent where the index 4 refers to gas upstream of the wave, index 3 to gas downstream of the
wave.The important point is that $u_{4}$ and $u_{3}$ are interpreted as velocities relative to the
wave; because it is stationary, they are in this case also relative to the laboratory.
 For the moving wave the velocities relative to the wave are $W$ (for the gas ahead)
and $W - u_{p}$ (for the gas behind) also making the total velocity equal to $W - u_{4}$ because
In figure, with the shock wave propagating to the right .
 After replacing $u_{4}$ and $u_{3}$, equation ~\ref{eq:9.1} to
~\ref{eq:9.3} become

\begin{equation} \label{eq: 9.4}
 \rho_{4}\left(W - u_{4}\right) = \rho_{3}\left(W - u_{p}\right)
\end{equation}

\begin{equation} \label{eq: 9.5}
 P_{4} + \rho_{4}\left(W - u_{4}\right)^{2} = P_{3} + \rho_{3}\left(W - u_{p}\right)^{2}
\end{equation}

\begin{equation} \label{eq: 9.6}
 h_{4} +\frac{\left(W - u_{4}\right)^{2}}{2} = h_{3} + \frac{\left(W - u_{p}\right)^{2}}{2}
\end{equation}

\noindent Equations ~\ref{eq: 9.4} to ~\ref{eq: 9.6} are the normal-shock equations for a shock moving with
velocity W into a stagnant gas. They can be rearranged and substituted, using h = e+p/$\rho$,
to obtain

\begin{equation} \label{eq: 9.7}
 e_{3} - e_{4} = \frac{P_{4}+P_{3}}{2}\left(V_{4}-V_{3}\right)
\end{equation}

\noindent The states after the shock are connected by the Rankine Hugoniot
shock jump conditions\cite{9} (Landau page no. 334) is valid, because we assume the case of polytropic gas,
$ e = C_{P}T$, $T = \frac{PV}{R}$ and $V = \frac{1}{\rho}$; equation ~\ref{eq: 9.7} becomes

\begin{equation} \label{eq: 9.8}
\boxed{ \frac{V_{3}}{V_{4}} = \frac{P_{3}}{P_{4}}\Bigg(\frac{\frac{\gamma + 1}{\gamma -1}
 + \frac{P_{3}}{P_{4}}}{1 + \frac{\gamma + 1}{\gamma -1}}\Bigg)}
\end{equation}

\begin{equation} \label{eq: 9.9}
\boxed{ \Rightarrow \frac{\rho_{3}}{\rho_{4}} = \frac{\frac{\gamma + 1}{\gamma -1}
 + \frac{P_{3}}{P_{4}}}{1 + \frac{\gamma + 1}{\gamma -1}}}
\end{equation}

\noindent Equations ~\ref{eq: 9.8} and ~\ref{eq: 9.9} give the temperature and
 density ratios across the shock wave as functions of the pressure 
ratio. We define  the moving shock Mach number as

\[Ma_{S} = \frac{W - u_{4}}{a_{4}}\]

\vspace{3 mm}

\noindent By taking the polytropic  gas relations and Eqn. ~\ref{eq: 9.4} to ~\ref{eq: 9.6} into account, we can
derive

\[\frac{P_{3}}{P_{4}} = 1 + \frac{2\gamma}{\gamma +1}\left(Ma_{S}^{2} - 1\right)\]

\begin{equation} \label{eq: 9.10}
 \Rightarrow Ma_{S} = \sqrt{\frac{\gamma +1}{2\gamma}\left(\frac{P_{3}}
{P_{4}} - 1\right) + 1}
\end{equation}
 
\noindent Since  $Ma_{S} = (W -u_{4})/a_{4}$, Eqn. ~\ref{eq: 9.10} leads to
\begin{equation} \label{eq: 9.11}
 W - u_{4} = a_{4}\sqrt{\frac{\gamma +1}{2\gamma}\left(\frac{P_{3}}
{P_{4}} - 1\right) + 1}
\end{equation}

\noindent Equation ~\ref{eq: 9.11} relates the velocity of the moving shock wave to the pressure ratio across
the wave and the local speed of sound of the gas ahead of the shock wave.

\noindent The last quantity we are interested in is the velocity up of the mass motion induced
by the shock wave. Equation ~\ref{eq: 9.4} can be rewritten

\begin{equation} \label{eq: 9.12}
 \boxed{u_{p} = W +  \left(W -u_{4}\right)\left(- \frac{\rho_{4}}{\rho_{3}}\right)}
\end{equation}

After substitution of Eqn. ~\ref{eq: 9.9} and ~\ref{eq: 9.11} into Eqn. ~\ref{eq: 9.12}, we obtain

\begin{equation} \label{eq: 9.13}
 \boxed{u_{p} = u_{4} + \frac{a_{4}}{\gamma}\left(\frac{P_{3}}{P_{4}}\right)
\left(\frac{\frac{2\gamma}{\gamma +1}}{\frac{P_{3}}{P_{4}} + \frac{\gamma - 1
}{\gamma +1}}\right)^{1/2}}
\end{equation}

\noindent With what we have derived so far, we can obtain for a given pressure ratio $P_{3}/P_{4}$ and
speed of sound $a_{4}$ the corresponding values of $\rho_{3}/\rho_{4}$, $T_{3}/T_{4}$, W and up from Eqn ~\ref{eq: 9.8},
~\ref{eq: 9.9}, ~\ref{eq: 9.11} and ~\ref{eq: 9.13}. The next section deals with the counterpart of the moving shock wave, namely the
expansion wave.

\vspace{5 mm}

\subsection{The incident expansion wave}

\noindent In the last section discussions about the 
relations between the regions 3 and 4 of Fig. ~\ref{fig: 6}. In this section we examine the regions
1 and 2. The formulas relate to a left-running expansions wave; they would be similar for a right-running
one except some changes of the signs in the equations.  We use the fact that the gas in region 1 feels as if a piston
 was pulled to the right with velocity $u_{4}$. In fact, $u_{4}$ is the
mass-motion velocity of the gas behind the expansion wave\cite{3} because this portion is high pressure region realtive
to region 4.

\vspace{3 mm}

\noindent Inside the wave, a mass motion with velocity u is induced, 
directed toward right. Temperature T and subsequently the local speed
 of sound a are reduced; because of this the head of the wave propagates 
faster into region 1 than other parts of the wave, so the
wave is spreading out while running down the tube.

\begin{figure}[h!]
 \centering
 \includegraphics[width=9cm,height=9cm,bb=0 0 426 361,keepaspectratio=true]{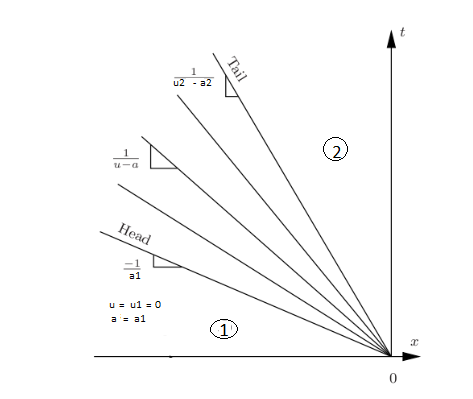}
 \caption{Characteristics for incident expansion wave centred at 0.}
 \label{fig:7}
\end{figure}

\noindent The tail of the wave moves at velocity $dx/dt = u_{2} -a_{2}$.
 Note that if $u_{2} > a_{2}$, i. e., $u_{2}$ is supersonic, the tail
 of the wave propagates to the right although the wave is left-running.
It can be shown that the characteristics of the expansion wave are 
straight lines and thus  the values of u, a (and consequently also p,
 $\rho$, T etc.) are constant along them. Such a wave is called a simple
 wave; a non simple waves where the characteristics are curved
comes up for example when the expansion wave is reflected.

\vspace{3 mm}

\noindent To obtain a closed analytical form for the expansion wave, we use Riemann invariant form \cite{9} (page no. 394)

\begin{equation} \label{eq: 9.14}
  u + \frac{2a}{\gamma -1 } = \text{constant through the wave}
\end{equation}

\noindent Applying this equation in region 1, we get

\begin{equation} \label{eq: 9.15}
 u_{1} + \frac{2a_{1}}{\gamma -1 }  = \text{constant}
\end{equation}

\noindent Combining these equations we get

\begin{equation} \label{eq: 9.16}
 \boxed{\frac{a}{a_{1}} = 1 - \frac{\gamma -1}{2}\left(\frac{u}{a_{1}} -\frac{u_{1}}{a_{1}}\right)}
\end{equation}

\noindent Which relates u and a within the expansion wave. Because $a = \sqrt{\gamma R T}$
Eqn ~\ref{eq: 9.16} also gives

\begin{equation} \label{eq: 9.17}
 \frac{T}{T_{1}} = \left(1 - \frac{\gamma -1}{2}\left(\frac{u}{a_{1}} - \frac{u_{1}}{a_{1}}\right)\right)^{2}
\end{equation}

\noindent For isentropic process $\triangle s = 0$

\[P = (\text{constant})\rho^{\gamma},\hspace{3mm} T = (\text{constant})\rho^{\gamma -1}
 \hspace{3mm}, a = (\text{constant})\rho^{(\gamma-1)/2}\]

\noindent Then from Eqn. ~\ref{eq: 9.17}, we get

\begin{equation} \label{eq: 9.18}
 \boxed{\frac{P}{P_{1}} = \left(1 - \frac{\gamma -1}{2}\left(\frac{u}{a_{1}} - \frac{u_{1}}{a_{1}}\right)\right)^{2\gamma/(\gamma -1)}}
\end{equation}
\noindent and

\begin{equation} \label{eq: 9.19}
 \boxed{\frac{\rho}{\rho_{1}} = \left(1 - \frac{\gamma -1}{2}\left(\frac{u}{a_{1}} - \frac{u_{1}}{a_{1}}\right)\right)^{2/(\gamma -1)}}
\end{equation}

\noindent These equations give all properties within the simple expansion 
wave as a function of the local gas velocity u. To get them as function of
x and t, we use  straight line characteristics as

\[\frac{dx}{dt} = \left(u - a\right)t\]

\[\Rightarrow x = \left(u - a\right)t + u_{1}\]

\noindent Using this relation in Eqn. ~\ref{eq: 9.16}

\[x = \left(u - u_{1} - a_{1} + \frac{\gamma -1}{2}u\right)t\]

\begin{equation} \label{eq: 9.20}
 \boxed{\Rightarrow u = u_{1} + \frac{2}{\gamma -1}\left(a_{1} + \frac{x}{t}\right)}
\end{equation}

\noindent The properties can be determined within the expansion wave, 
$u_{1} - a_{1} \leqslant x/t \leqslant u_{2} - a_{2}$

\subsection{Final Expression of Shock Tube}

\noindent Finally, we combine the last two section for a closed analytical
solution. End of the day we recall that $u_{2} = u_{3} = u_{p}$ and
$P_{2} = P_{3}$ across the contact surface; last expression of $u_{p}$ was

\[u_{p} = u_{3} = u_{4}  + \frac{a_{4}}{\gamma}\left(\frac{P_{3}}{P_{4}}\right)
\left(\frac{\frac{2\gamma}{\gamma +1}}{\frac{P_{3}}{P_{4}} + \frac{\gamma - 1
}{\gamma +1}}\right)^{1/2}\]

\noindent Applying Eqn. ~\ref{eq: 9.18} to the tail of the expansion wave,

\begin{equation} \label{eq: 9.21}
 \frac{P_{2}}{P_{1}} = \left(1 - \frac{\gamma- 1}{2}\Big(\frac{u_{2}}{a_{1}}-\frac{u_{1}}{a_{1}}\Big)\right)^{2\gamma/(\gamma - 1)}
\end{equation}
\[\Rightarrow u_{2} = u_{1} + \frac{2a_{1}}{\gamma -1}\left(1 - \Big(\frac{P_{2}}{P_{1}}\Big)^{(\gamma - 1)/2\gamma}\right)\]

\noindent Putting $P_{2} = P_{3}$

\[u_{2} = u_{1} + \frac{2a_{1}}{\gamma -1}\left(1 - \Big(\frac{P_{3}}{P_{1}}\Big)^{(\gamma - 1)/2\gamma}\right)\]

\noindent Since $u_{2} = u_{3}$, so we apply in last Eqn. and equation ~\ref{eq: 9.13}, we get

\[u_{4} + \frac{a_{4}}{\gamma}\bigg(\frac{P_3}{P_{4}}\bigg)\Bigg(\frac{\frac{2\gamma}{\gamma +1}}{\frac{P_{3}}{P_{4}} + \frac{\gamma - 1
}{\gamma +1}}\Bigg)^{1/2} = u_{1} + \frac{2a_{1}}{\gamma -1}\left(1 - \Big(\frac{P_{3}}{P_{1}}\Big)^{(\gamma - 1)/2\gamma}\right)\]

\vspace{3 mm}
\begin{equation} \label{eq: 9.22}
 \boxed{ \frac{P_{4}}{P_{1}} = \frac{P_{4}}{P_{3}}\left(1 + \frac{\big(u_{1} - u_{4}\big)\big(\gamma -1\big)}{2a_{1}} -\frac{\Big(\gamma -1\Big)
\Big(\frac{a_{4}}{a_{1}}\Big)\Big(\frac{P_{3}}{P_{4}} -1\Big)}{\sqrt{2\gamma\Bigg(\Big(\gamma -1\Big) + \Big(\gamma +1\Big)\frac{P_{3}}{P_{4}}\Bigg)}}\right)^{2\gamma/(\gamma -1)}}
\end{equation}

\noindent Equation (9.22) gives the incident shock strength $P_{3}/P_{4}$ as an implicit function of the
diaphragm pressure ratio $P_{1}/P_{4}$. We can now unfold a recipe for the solution of the shock
tube problem, which consists of all the boxed equations:

\begin{enumerate}
 \item[$\ast$] Calculate $P_{3}/P_{4}$ from Eqn. ~\ref{eq: 9.22} to get the strength of the shock wave.
\item[$\ast$] Calculate all other shock properties from Eqn. ~\ref{eq: 9.8}, ~\ref{eq: 9.9}, ~\ref{eq: 9.11} and ~\ref{eq: 9.13}.
\item[$\ast$] Calculate $P_{2}/P_{1} = (P_{2}/P_{4})/(P_{1}/P_{4}) = (P_{3}/P_{4})/(P_{1}/P_{4})$ to get the strength of the expansion wave.
\item[$\ast$] All other properties behind the expansion wave can be found using
\[\frac{P_{2}}{P_{1}} = \left(\frac{\rho_{2}}{\rho_{1}}\right)^{\gamma}\]
\item[$\ast$] The properties inside the expansion wave can be found from Eqs. ~\ref{eq: 9.16} to ~\ref{eq: 9.19} and ~\ref{eq: 9.20}.
\end{enumerate}

\subsection{Implementation in Python}

\noindent Source code of these mathematics is given in section ~\ref{subsec:Code For Sod Shock Tube Problem By Python Code:}, where xgrid is a vector
containing the x-positions where the solution is evaluated rho , u ,c(=a),P,machE  and $entropy\_E$ are vectors
of the size of x\_grid containing the density, local velocity,sound velocity,pressure, Mach number and entropy at the
corresponding positions.

\vspace{3mm}

\noindent $t\_end$ specifies at what time the solution is to be evaluated;rhoL($=\rho_{1}$), 
rhoR $( = \rho_{4})$ , PrL $(P_{1})$, PrR $(= P_{4})$, uL = $u_{1}$ uR = $u_{4}$ initial
settings for density, pressure and local velocity  in the two compartments.

\vspace{3 mm}

\noindent After parsing the input and optionally setting default values, some constants are specified:
the heat capacity gamma = 1.4 and position of diaphragm at 0.5 . Next local speed 
and the densities for region 1 and  4 are computed from the quantities
known so far. To get P3, rho3, rho2, u2 and c2 we solve Eqn ~\ref{eq: 9.22} for $P_{3}/P_{4}$
that means we have to find a $P_{3}/P_{4}$ satisfying

\begin{equation}
  \frac{P_{4}}{P_{3}}\left(1 + \frac{\big(u_{1} - u_{4}\big)\big(\gamma -1\big)}{2a_{1}} -\frac{\Big(\gamma -1\Big)
\Big(\frac{a_{4}}{a_{1}}\Big)\Big(\frac{P_{3}}{P_{4}} -1\Big)}{\sqrt{2\gamma\Bigg(\Big(\gamma -1\Big)
 + \Big(\gamma +1\Big)\frac{P_{3}}{P_{4}}\Bigg)}}\right)^{2\gamma/(\gamma -1)} - \frac{P_{4}}{P_{1}} = 0
\end{equation}

\noindent This equation is solved by python function $\textbf{fsolve}$ which can gives the iterative solution about a point
of iterative solution by using Newton's method, bisection method etc.

\noindent  Now, the mesh $x\_grid$ is initialized, usually to a size of 1 x 300, but the number of
points could also be changed to some other value if desired. Before iterating through all
the solution vectors, the boundaries of the regions are determined using the knowledge
about the velocities of head and tail of the expansion wave.

\begin{enumerate}
 \item[$\ast$] $\textbf{Starting expansion fan}$ i.e Boundary between leftmost driver gas and expansion wave. $\hspace{10cm}$
\[pos1 = 0.5 + (u_{1} - a_{1}) * t\_end \]
\[\Rightarrow pos1 = 0.5 + (uL - aL) * t\_end \]
\item[$\ast$] $\textbf{End of expansion fan}$ i.e. Boundary between expansion wave and lower pressure driver gas.$\hspace{10cm}$
\[pos2 = 0.5 + (u_{3} - a_{3} ) * t\_end \]
\[\Rightarrow pos2 = 0.5 + (u2 + uR - c2) * t\_end \hspace{3mm}\text{, because u2 = u3}\]
\item[$\ast$] $\textbf{Position of contact discontinuity}$ i.e. Boundary between driver gas and driven gas.$\hspace{10cm}$
\[conpos = 0.5 + u\_p*t\_end \]
\[\Rightarrow conpos = 0.5 + (u2 + uR)*t\_end \]
\item[$\ast$] $\textbf{Shock Position}$ i.e Location of the shock wave.$\hspace{10cm}$
\[spos = 0.5  + t\_end* W\]
\[\Rightarrow spos = 0.5 +  t\_end* \left(a_{4}\sqrt{\frac{\gamma +1}{2\gamma}\left(\frac{P_{3}}{P_{4}}-1\right) +1} + u_{4}\right)\]
\[\Rightarrow spos = 0.5 + t\_end *\left(a_{4}\sqrt{\frac{\gamma -1}{2\gamma} + \frac{\gamma +1 }{2\gamma}\frac{P_{3}}{P_{4}}} + u_{4}\right) \]
\[\Rightarrow spos = 0.5 + t\_end * \left(cR\sqrt{\frac{\gamma -1}{2\gamma} + \frac{\gamma +1 }{2\gamma}\frac{P3}{P4}} + uR\right) \]
\end{enumerate}

\vspace{10 mm}

\section{Sod Shock Tube}

\noindent  The Sod shock tube problem, named after Gary A. Sod, is a common test 
for the accuracy of computational fluid codes, like Riemann solvers and
 was heavily investigated by Sod in 1978. The test consists of a one 
dimensional Riemann problem with the following parameters, for left and
 right states of an ideal gas. 

\begin{equation}
 \begin{pmatrix}
  \rho_{L} \\
u_{L} \\
P_{L}
 \end{pmatrix}
=\begin{pmatrix}
  1.0 \\
0.0 \\
1.0
 \end{pmatrix}
\begin{pmatrix}
      \rho_{R}\\
v_{R} \\
P_{R}
\end{pmatrix}
=
\begin{pmatrix}
 0.125 \\
0.0 \\
0.1
\end{pmatrix}
\end{equation}
\noindent  The time evolution of this problem can be described by solving the Euler
 equations. Which leads to three characteristics, describing the propagation
 speed of the various regions of the system. Namely the rarefaction wave, the
 contact discontinuity and the shock discontinuity. If this is solved numerically,
 one can test against the analytical solution, and get information how well
 a code captures and resolves shocks and contact discontinuities and reproduce
 the correct density profile of the rarefaction wave.

\section{Result:}
\subsection{Shock tube Data Plot after Breaking the Diaphragm}
\noindent In this case

\begin{table}[h]
\begin{centering}

\begin{tabular}{| c || c | c |}
\hline
 Quantity & left side & right side\\
\hline
\hline
Density & 1.0 & 0.125\\
\hline
Pressure & 1.0 & 0.1\\
\hline
Velocity & 0.0 & 0.0\\
\hline
\multicolumn{3}{ |c| }{ $t\_end$=0.17 Sec   N=300   $\mu $=0.35} \\
\hline
\end{tabular}
\caption{Shock tube data at t = 0}
\label{Table: 1}
\end{centering}
\end{table}

\begin{figure}[h!]
 \centering
 \includegraphics[width=14cm,height=16cm,keepaspectratio=true]{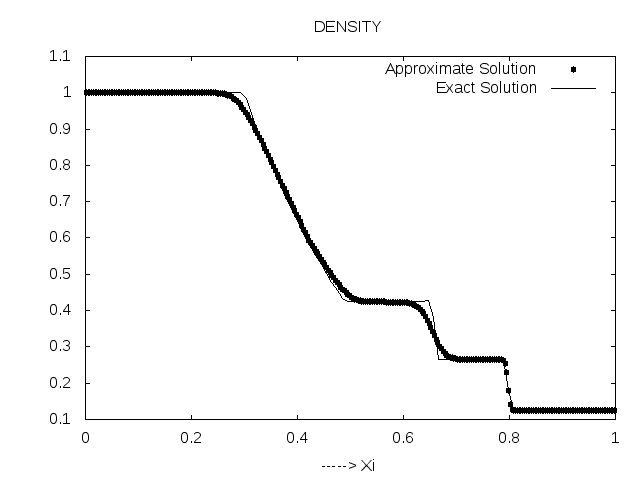}
 \caption{Density Distribution}
 \label{fig:8}
\end{figure}

\newpage

\begin{figure}[h!]
 \centering
 \includegraphics[width=13.7cm,height=16cm,bb=0 0 640 480,keepaspectratio=true]{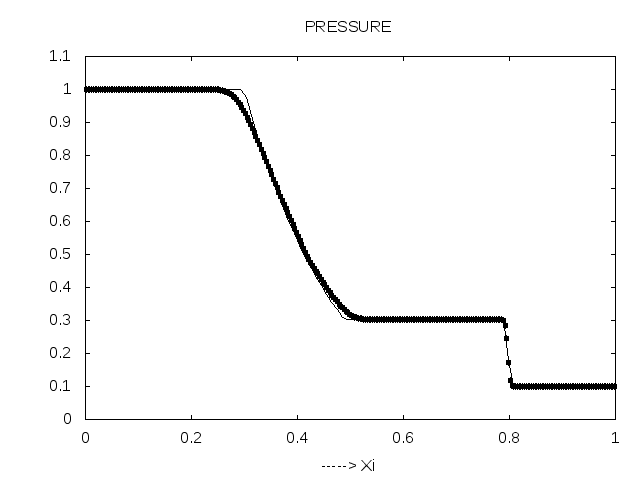}
\caption{Pressure Distribution}
 \label{fig:8}
\end{figure}

\begin{figure}[h!]
 \centering
 \includegraphics[width=13.7cm,height=16cm,bb=0 0 640 480,keepaspectratio=true]{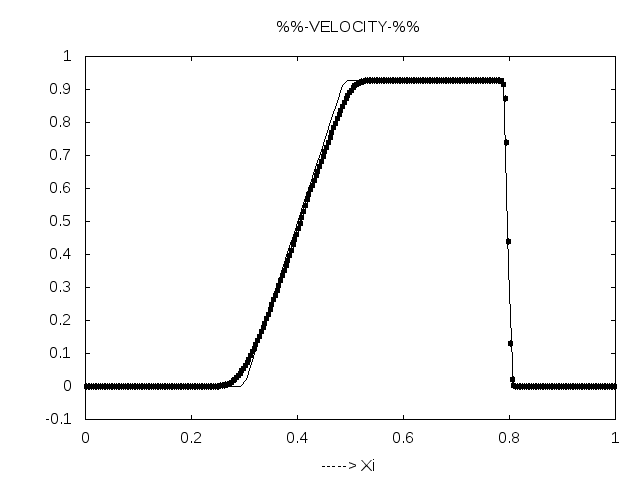}
\caption{Velocity Distribution}
 \label{fig:9}
\end{figure}
\vspace{6mm}

\begin{figure}[ht]
 \centering
 \includegraphics[width=13.7cm,height=16cm,bb=0 0 640 480,keepaspectratio=true]{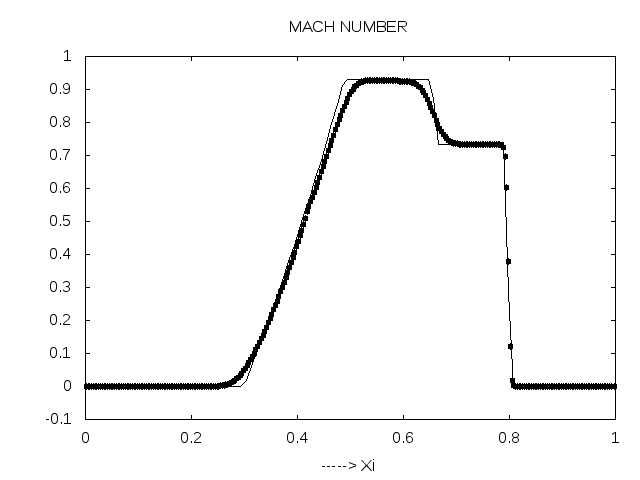}
\caption{Mach Number Distribution}
 \label{fig:9}
\end{figure}

\begin{figure}[h!]
 \centering
 \includegraphics[width=13.7cm,height=15cm,bb=0 0 640 480,keepaspectratio=true]{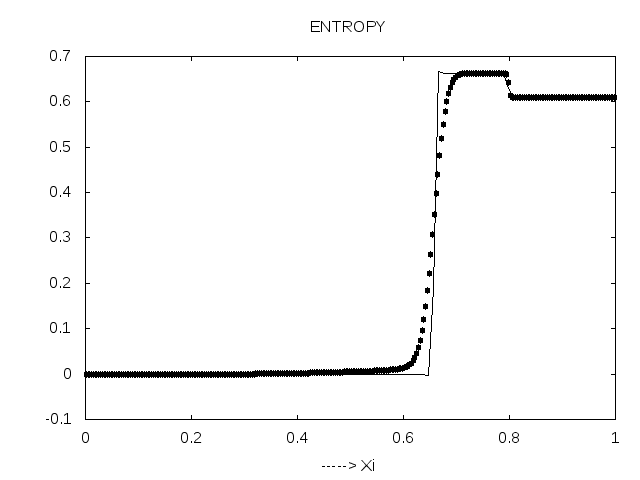}
\caption{Entropy Distribution}
 \label{fig:10}
\end{figure}

\noindent Comparison to measurements yields respectable results,
actually this result is only for stationary shock tube.but our program simulation
for both cases. 
\noindent Three Python scripts on the accompanying movies, which visualize the
density,pressure,velocity, Mach number and entropy in shock tube
after breaking the diaphragm.See Appendix ~\ref{subsec: Contents in CD} for details on
the CD.

\begin{table}[h]
\centering
\begin{tabular}{|l|c|c|}
\hline
Script file & Video file & Quantity\\
\hline
\hline
exact.py & exact.avi & Exact value by Shock tube formula\\
\hline
approximate.py & approximate.avi & Approximate solution by Roe Solver\\
\hline
approximate\_exact.py & approximate\_exact.avi & Both solution\\
\hline

\end{tabular}
\caption{Scripts for video files visualizing the analytical and approximate solution}
\label{tab:template}
\end{table}
\section{Conclusions}
We described a strategy for obtaining numerical solutions to hyperbolic
initial-value problems by Roe Scheme Method\cite{1}. In the project we have the details of one
element in that strategy. Our investigations into the Euler equations have revealed 
some very tidy structures. Also the existence of a fast exact Riemann solver is 
found more generally useful. Our programming experience is that the present direct
method is about a time-consuming as one cycle of the iterative procedures. 
Iterations needed to obtain  more exact solutions.
Also we use Harten entropy fix assuming the Harten original approach: the entropy fix
as means to select the physically relevant  weak solution, corresponding to the vanishing
viscosity solution.

\newpage

\section{Appendix}
\label{sec:Appendix}

\subsection{Implementation in Python of Roe Scheme:}

\noindent we will apply the Roe Scheme to the one test case of shock tube.
We discretize the flow of quantity $\textbf{u}$ by finite volume method of
equation ~\ref{eq:7.7} in time by explicit Euler method

\[\bar u_{i}(t_{2}) = \bar u_{i}(t_{1}) - \frac{\triangle t}{\triangle x_{i}}[F_{i+1/2} - F_{i-1/2}]\]
\begin{equation} \label{eq: 13.1}
 \Rightarrow \bar u_{i}(t_{2}) = \bar u_{i}(t_{1}) - \nu [F_{i+1/2} - F_{i-1/2}]
\end{equation}

\noindent The exact solution is used to prescribe $u_{1}(t)$ and $u_{i}(t)$. The time step 
$\triangle t$ and mesh size $\triangle x_{i}$'s are constant .

\begin{enumerate}
 \item[$\ast$]$\textbf{Grid of Physical Quantity}$ Take x\_grid, density grid, pressure grid, velocity grid all have same no of grid point(in our case 300)
 last three grid maintain Riemann Problem given in equation ~\ref{eq: 8.2} i.e. first half grid value will rhoL, PrL, uL and last half grid value rhoR, PrR, uR.
\item[$\ast$] $\textbf{Energy Momentum}$ Calculate total energy, momentum, total enthalpy by using equation ~\ref{eq: 8.8} and ~\ref{eq: 8.9}
\item[$\ast$]$\textbf{Average Calculation}$ Calculate density average velocity, enthalpy average by using equation ~\ref{eq: 8.18} and ~\ref{eq: 8.19} similarly 
calculate the flux averages and apply on all grid point.
\item[$\ast$] $\textbf{Grid of New Average Calculation}$ From this new grid calculate Aav by using equation ~\ref{eq: 8.11} also alpha1,Eqn ~\ref{eq: 8.25} alpha2, ~\ref{eq: 8.24} alpha3 ~\ref{eq: 8.26}, lambda's 
~\ref{eq: 8.21} now we have all the eigenvalues and corresponding eigenvector but to remove the problem of artificial viscosity apply Harten sonic entropy
fix  condition ~\ref{eq: 8.28}
\item[$\ast$] $\textbf{Flux Average}$ Calculate roe flux's Eqn ~\ref{eq: 8.27} where for 1st term is flux averages.
\item[$\ast$]$\textbf{Average data for grid formation}$ since we have the average data between two grid point so we can fill centering grid point by using ~\ref{eq: 13.1} , which
is our new grid.
\item[$\ast$] $\textbf{Entropy}$ End of the program calculate entropy
\[entropy = log(P/\rho **\gamma)\]
\end{enumerate}

\noindent Since at t = 0 all left and right side data are same so when we calculate the average value gives us '-nan' so next time gives two '-nan' data so it will be dangerous. 
To remove this problem if $U_{i} == -nan$ then $U_{i} = U_{i-1}$ .
\subsection{Python Code}

\lstset{
language=Python,
basicstyle=\ttfamily,
otherkeywords={self},             
keywordstyle=\ttfamily\color{orange!100!black},
keywords=[2]{func},
keywords=[3]{ttk},
keywords=[4]{range,len,int,str,True,False,file,open},
keywordstyle={[2]\ttfamily\color{blue!100!green}},
keywordstyle={[3]\ttfamily\color{red!100!orange}},
keywordstyle={[4]\ttfamily\color{violet!100!black}},
emph={MyClass,__init__},          
emphstyle=\ttfamily\color{red!100!black},    
stringstyle=\color{green!100!black},
showstringspaces=false            
}

\begin{lstlisting}
#!/usr/bin/env python
"""
Roe Scheme for one-dimensional Euler equations of Perfect
Gas in Adiabatic Process....
"""
import numpy as np
from pylab import *
from decimal import *
""" -----o---Specification of Riemann problems---o--- """

PrL = 1.0;  PrR = 0.1; rhoL = 1.0;  rhoR = 0.125;
uL = 0;  uR = 0; t_end = 0.17; mu = 0.35;	# nu = dt/dx

""" -------- Input Section -------- """
gamma = 1.4              # Ratio of specific heat
N = 300.0                 # Number of grid cells
"""-------END I/P ----- """
gammaA = gamma - 1 ; gammaB = 1/(gamma - 1)
dx = 1/N                   # Cell Size
dt = mu*dx                 # Time step
n = int(t_end/dt)
"""
               Definition of grid numbering
       x=0                                      x=1 
 grid    |--o--|--o--|--o--|--o--  .... ---|--o--|
            1  1  2  2  3  3  4           N-1 N  I
"""
N = int(N)
x = np.linspace(0.0,1.0,N+1)
xgrid = np.linspace(dx,1.0,N) - (dx/2)
Pr = np.zeros((1,len(xgrid)))
rho = np.zeros((1,len(xgrid)))
u = np.zeros((1,len(xgrid)))
for i in range (0, N, 1):
    if xgrid[i] < 0.5 : Pr[0,i] = PrL; rho[0,i] = rhoL; u[0,i] = uL
    else: Pr[0,i] = PrR; rho[0,i] = rhoR; u[0,i] = uR

eI = (gammaB*Pr)/rho             # specific (= per unit mass) internal energy
eT = gammaB*Pr                   # Thermal energy
M = u*rho                        # Momntum rho*u
Et = eT + 0.5*M* u               # Total energy
EtL = Et[0,0] ; EtR = Et[0,N-1]
Ht = gamma*eI + 0.5*u*u          # Total enthalpy
# New cell centre Variables
mach = np.zeros((1,N))
rho_n = np.zeros((1,N))
M_n = np.zeros((1,N))
Et_n = np.zeros((1,N))
A_n = np.zeros((1,N))
# Roe Fluxes
roeF1 = np.zeros((1,N))
roeF2 = np.zeros((1,N))
roeF3 = np.zeros((1,N))
# Roe Averages
Hav = np.zeros((1,N))
uav = np.zeros((1,N))
Aav = np.zeros((1,N))
dd = np.zeros((1,N))
# Flux Averages
F1av = np.zeros((1,N))
F2av = np.zeros((1,N))
F3av = np.zeros((1,N))
# State Vector Difference
delrho = np.zeros((1,N))
delM = np.zeros((1,N))
delEt = np.zeros((1,N))
# Eigenvalues
lamda1 = np.zeros((1,N))
lamda2 = np.zeros((1,N))
lamda3 = np.zeros((1,N))
#Coefficents
alpha1 = np.zeros((1,N))
alpha2 = np.zeros((1,N))
alpha3 = np.zeros((1,N))
t = 0
for i in range (0,n,1):
    t = t + dt    
    for j in range(0,N-2,1):
        dd[0,j] = np.sqrt(rho[0,j+1]/rho[0,j])
        if np.isnan(dd[0,j]) == True:
            dd[0,j] = dd[0,j-1]
        Hav[0,j] = (Ht[0,j] + dd[0,j]*Ht[0,j+1])/(1+dd[0,j])
        uav[0,j] = (u[0,j] + dd[0,j]*u[0,j+1])/(1+dd[0,j])
        delrho[0,j] = rho[0,j+1] - rho[0,j]
        delM[0,j] = M[0,j+1] - M[0,j]
        delEt[0,j] = Et[0,j+1] - Et[0,j]

        F1av[0,j] = 0.5*(M[0,j] + M[0,j+1])
        F2av[0,j] = 0.5*((M[0,j]*M[0,j]/rho[0,j]) + Pr[0,j] +
                       (M[0,j+1]*M[0,j+1]/rho[0,j+1]) + Pr[0,j+1])
        F3av[0,j] = 0.5*((M[0,j]*Ht[0,j]) + (M[0,j+1]*Ht[0,j+1]))
    
    Aav = np.sqrt(gammaA*(Hav - 0.5*(uav*uav)))
    
    Mav = uav/Aav
    alpha1 = (0.25*Mav*(2+gammaA*Mav)*delrho -
            0.5*(1+gammaA*Mav)*(delM/Aav) + 0.5*gammaA*(delEt/(Aav**2)))
    alpha2 = ((1-0.5*gammaA*(Mav**2))*delrho  + gammaA*(Mav/Aav)*delM -
              gammaA*(delEt/(Aav**2)))
    alpha3 = (-0.25*Mav*(2-gammaA*Mav)*delrho + 0.5*(1-gammaA*Mav)*(delM/Aav)
              + 0.5*gammaA*(delEt/(Aav**2)))
    lamda1 = np.abs(uav - Aav); lamda2 = np.abs(uav);
    lamda3 = np.abs(uav + Aav)
    """ Harten's sonic entropy fix """
    Eps = 0.5          
    """ Eps = 0.0	        No entropy fix """
    for j in range (0,N-1):
        if lamda1[0,j] < Eps :
            lamda1[0,j] = 0.5*((Eps + lamda1[0,j]**2)/Eps)
        if lamda3[0,j] < Eps:
            lamda3[0,j] = 0.5*((Eps + lamda3[0,j]**2)/Eps)

    roeF1 = F1av - 0.5*lamda1*alpha1 - 0.5*lamda2*alpha2 - 0.5*lamda3*alpha3
    roeF2 = (F2av - 0.5*lamda1*alpha1*(uav-Aav) - 0.5*lamda2*alpha2*uav
             - 0.5*lamda3*alpha3*(uav+Aav))
    roeF3 = (F3av - 0.5*lamda1*alpha1*(Hav - uav*Aav) - 0.25*lamda2*
	     alpha2*uav*uav- 0.5*lamda3*alpha3*(Hav + uav*Aav))
    rho_n[0,0] = rhoL; rho_n[0,N-1] = rhoR
    M_n[0,0] = rhoL*uL; M_n[0,N-1] = rhoR*uR
    Et_n[0,0] = EtL; Et_n[0,N-1] = EtR
    for  j in range(1, N-1):
        rho_n[0,j] = rho[0,j] - mu*(roeF1[0,j] - roeF1[0,j-1])
        M_n[0,j] = M[0,j] -mu*(roeF2[0,j] - roeF2[0,j-1])
        Et_n[0,j] = Et[0,j] - mu*(roeF3[0,j] - roeF3[0,j-1])        
    u = M_n/rho_n
    Pr = gammaA*(Et_n - 0.5*M_n*u)
    eI = (gammaB*Pr)/rho_n
    Ht = (Et_n + Pr)/rho_n
    rho = rho_n
    M = M_n
    Et = Et_n
    A_n = np.sqrt(gamma*gammaA*eI)
    mach = u/A_n  
    for k in range(1,N):
        if np.isnan(u[0,k]) == True:
            u[0,k] = u[0,k-1]
        if np.isnan(Pr[0,k]) == True:
            Pr[0,k] = Pr[0,k-1]
        if np.isnan(eI[0,k]) == True:
            eI[0,k] = eI[0,k-1]
        if np.isnan(Ht[0,k]) == True:
            Ht[0,k] = Ht[0,k-1]
        if np.isnan(rho[0,k]) == True:
            rho[0,k] = rho[0,k-1]
        if np.isnan(M[0,k]) == True:
            M[0,k] = M[0,k-1]
        if np.isnan(Et[0,k]) == True:
            Et[0,k] = Et[0,k-1]
ent_ropy = np.log(Pr/(rho**gamma))

""" ---o---File Write Section---o--- """
file = open("roe_scheme.txt","w")
for i in range(0,xgrid.size):
    (file.write(str("%.6f"%xgrid[0,i])+'   '+str("%.6f"%rho_n[0,i])+'
      '+str("%.6f"%Pr_n[0,i])+'  '+str("%.6f"%u_nE[0,i])+'
      '+str("%.6f"%machE[0,i])+'    '+str("%.6f"%entropy_E[0,i])))
    file.write("\n")
file.close()
\end{lstlisting}

\subsection{Code For Sod Shock Tube Problem By Python Code:}

\label{subsec:Code For Sod Shock Tube Problem By Python Code:}

\lstset{
language=Python,
basicstyle=\ttfamily,
otherkeywords={self,as},             
keywordstyle=\ttfamily\color{orange!100!black},
keywords=[2]{func},
keywords=[3]{ttk},
keywords=[4]{range,len,int,str,True,False,file,open},
keywordstyle={[2]\ttfamily\color{blue!100!black}},
keywordstyle={[3]\ttfamily\color{red!100!orange}},
keywordstyle={[4]\ttfamily\color{violet!100!blue}},
emph={MyClass,__init__},          
emphstyle=\ttfamily\color{red!100!black},    
stringstyle=\color{green!100!black},
showstringspaces=false            
}

\begin{lstlisting}
#!/usr/bin/env python

import numpy as np
from scipy.optimize import fsolve

#--------o--Input Section--o-------
# Sod Shock Tube Problem
PrL = 1.0;  PrR = 0.1; rhoL = 1.0;  rhoR = 0.125;
uL = 0.0;  uR = 0.0; t_end = 0.2; mu = 0.35;    # mu = dt/dx
#--------o--END Input Section--o-----------------      
gamma = 1.4
gammaA = gamma - 1.0
gammaB = 1/gammaA
gammaC = gamma + 1.0

PRL = PrR/PrL
cR = np.sqrt(gamma * PrR/rhoR)
cL = np.sqrt(gamma * PrL/rhoL)
CRL = cR/cL
machL = (uL - uR)/cL
def func(p34):
        wortel = np.sqrt(2 * gamma * (gammaA + gammaC * p34))
        yy = (gammaA * CRL * (p34 - 1)) / wortel
        yy = (1 + machL * gammaA/2 - yy)**(2 * gamma/gammaA)
        y = yy/p34 - PRL
        return y

p34 = fsolve(func,3.0)          # p34 = p3/p4
print p34
p3 = p34 * PrR
alpha = gammaC/gammaA
rho3 = rhoR * (1 + alpha * p34)/(alpha + p34)
rho2 = rhoL * (p34 * PrR/PrL)**(1/gamma)
u2 = uL-uR+(2/gammaA)*cL*(1 - (p34 * PrR/PrL)**(gammaA/(2 * gamma)))
c2 = np.sqrt(gamma * p3/rho2)
spos = (0.5 + t_end* cR *np.sqrt(gammaA/(2 * gamma) +
        gammaC/(2 * gamma) * p34) + t_end * uR)  # Shock position

conpos = 0.5 + u2 * t_end + t_end * uR   # Position of contact discontinuity 
pos1 = 0.5 + (uL - cL) * t_end           # Start of expansion fan
pos2 = 0.5 + (u2 + uR - c2) * t_end      # End of expansion fan

xgrid = np.linspace(0,1,100)
PrE = np.zeros((1,len(xgrid)))
uE= np.zeros((1,len(xgrid)))
rhoE = np.zeros((1,len(xgrid)))
machE = np.zeros((1,len(xgrid)))
cE = np.zeros((1,len(xgrid)))
xgrid = np.matrix(xgrid)

for i in range (0,xgrid.size):   
        if xgrid[0,i] <= pos1:
                PrE[0,i] = PrL
		rhoE[0,i] = rhoL
                uE[0,i] = uL
		cE[0,i]= np.sqrt(gamma*PrE[0,i]/rhoE[0,i]);
                machE[0,i]= uE[0,i]/cE[0,i];
        elif xgrid[0,i] <= pos2:
                PrE[0,i] = (PrL*(1 + (pos1 - xgrid[0,i])
		  /(cL * alpha * t_end))**(2 * gamma/gammaA))
                rhoE[0,i] = (rhoL*(1+(pos1 - xgrid[0,i])
		      /(cL * alpha * t_end))**(2/gammaA))
                uE[0,i] = uL + (2/gammaC)*(xgrid[0,i] - pos1)/t_end
                cE[0,i] = np.sqrt(gamma * PrE[0,i]/rhoE[0,i])
                machE[0,i] = uE[0,i]/cE[0,i]
        elif xgrid[0,i] <= conpos:
                PrE[0,i] = p3
		rhoE[0,i] = rho2
                uE[0,i] = u2 + uR;    
		cE[0,i]= np.sqrt(gamma * PrE[0,i]/rhoE[0,i]);
                machE[0,i] = uE[0,i]/cE[0,i]
        elif xgrid[0,i] <= spos:
                PrE[0,i] = p3
		rhoE[0,i] = rho3
		uE[0,i] = u2+uR
                cE[0,i] = np.sqrt(gamma * PrE[0,i]/rhoE[0,i])
                machE[0,i] = uE[0,i]/cE[0,i]
        else:
                PrE[0,i] = PrR
		rhoE[0,i] = rhoR;
                uE[0,i] = uR
		cE[0,i] = np.sqrt(gamma*PrE[0,i]/rhoE[0,i]);
                machE[0,i] = uE[0,i]/cE[0,i];
 entropy_E = np.log(PrE/rhoE**gamma);

""" ---o---File Write Section---o--- """

file = open("Riemann.txt","w")
for i in range(0,xgrid.size):
    (file.write("%.6f"%str(xgrid[0,i])+'   '+str("%.6f"%rhoE[0,i])+'
  '+str("%.6f"%PrE[0,i])+'  '+str("%.6f"%uE[0,i])+'
   '+str("%.6f"%machE[0,i])+'	'+str("%.6f"%entropy_E[0,i])))
    file.write("\n")
file.close()


\end{lstlisting}

%
%
%

%
%


\begin{thebibliography}{99}
\bibitem{1}\href{http://www.sciencedirect.com/science/article/pii/0021999181901285}{P. L. ROE JOURNAL OF COMPUTATIONAL. PHYSICS. 43, 357-372 (1981). Approximate.
 Riemann Solvers, Parameter Vectors, and Difference Schemes.}

\bibitem{2} S. K. GHOSH `LECTURE ON ATMOSPHERIC SCIENCE. Fluid Mechanics` 


\bibitem{3}\href{http://cdn.preterhuman.net/texts/science_and_technology/physics/Fluid_Mechanics/Computational%20Fluid%20Mechanics%20And%20Heat%20Transfer%20-%20Anderson.pdf}
{Anderson, Dale A., Tannehill, John C., \& Pletcher, Richard H.
 1984. Computational Fluid Mechanics and Heat Transfer. New York: McGraw-Hill.}
\bibitem{4} \href{http://www.icm.csic.es/scimar/pdf/61/sm61s1007.pdf}{T.J. PEDLEY, 'Introduction to Fluid Dynamics', SCI. MAR., \textbf{61} (Supl. 1): 7-24}
\bibitem{5} \href{http://en.wikipedia.org/wiki/Finite-volume_method}{ R. Eymard, T Gallouët and R. Herbin, 'FINITE VOLUME METHOD', update of the article published in Handbook of Numerical Analysis, 2000}
\bibitem{6} \href{http://www.eos.ubc.ca/~ehearn/EOSC_453/Reynolds_Transport.pdf}{Elizabeth H. Hearn,'The Reynolds Transport Theorem'}
\bibitem{7} \href{http://books.google.co.in/books?id=VwwTHyB8WKcC&pg=PA292&lpg=PA292&dq=P.+Wesseling,+Principles+of+Computational+Fluid+Dynamics&source=bl&ots=4eoOIyK_7r&sig=4SZDNcArhcLSemVLuX9njTt1cEU&hl=en&sa=X&ei=UyJ_UfCzBMXyrQfrsYH4Dw&ved=0CDkQ6AEwAg}{P. Wesseling, Principles of Computational Fluid Dynamics}
\bibitem{8} \href{http://webcache.googleusercontent.com/search?q=cache:leS5kPO8Bf4J:www.aero.polimi.it/~quartape/bacheca/materiale_didattico/ef_JCP.ps+&cd=1&hl=en&ct=clnk&gl=in}{M. Pelanti, L. Quartapelle and L. Vigevano 'A review of entropy fixes as applied to Roe's linerization'}
\bibitem{9} {L. D. LANDAU \& E. M. LIFSHITZ `FLUID MECHANICS' Second Edition }
\end{thebibliography}
\end{document}